\documentstyle[12pt]{article}
\begin{document}

\title{Late-Type Dwarf 
Galaxies in the Virgo Cluster: II. Star Formation Properties}
\author{Elchanan Almoznino\thanks{email - nan@wise.tau.ac.il}
 \and Noah Brosch}
\date{{\it Wise Observatory \& School of Physics
and Astronomy,\\ Raymond and Beverly Sackler Faculty of Exact Sciences,\\
Tel-Aviv University}}
\maketitle
 
\begin{abstract}

We study 
star-formation-inducing mechanisms in galaxies
through multi-wavelength measurements of a sample
 of dwarf galaxies in the 
Virgo cluster described in paper I.
Our main goal is to test how star formation
inducing mechanisms depend on several
parameters of the galaxies, such as morphological type and hydrogen content.
We derive the star formation rate and star formation histories
of the galaxies, and check their dependence on other parameters.

Comparison of the sample galaxies with population synthesis models
shows that these objects have significantly lower metallicity than the
Solar value. The colors can generally
be explained as a combination of two different stellar populations: a young
(3--20 Myr) metal-poor population which represents the stars currently forming
presumably in a starburst, and
an older (0.1--1 Gyr) population of previous stellar generations. 
There is evidence that the older stellar population was also formed in a starburst.
This is consistent with the
explanation that star formation in this type of objects takes
place in short bursts followed by long quiescent periods.

No significant correlation is found between the
star formation properties of the sample galaxies and their hydrogen content.
Apparently, when star formation occurs in bursts, other parameters influence 
the star formation properties more significantly than the amount of atomic
hydrogen. No correlation is found between the projected
Virgocentric distance and the rate of star formation in the galaxies, suggesting
that tidal interactions are not significant in triggering star formation in
cluster dwarf galaxies.

\end{abstract}


    \section{Introduction} \label{sec_int}

Star formation is probably the most fundamental 
process in galaxies. Understanding its nature, 
and its dependence
on galactic type and environment may contribute to our knowledge
about development of galaxies, as well as about the development of the entire 
Universe.

\begin{figure}[htbp]
\vspace{9.4cm}
\includegraphics{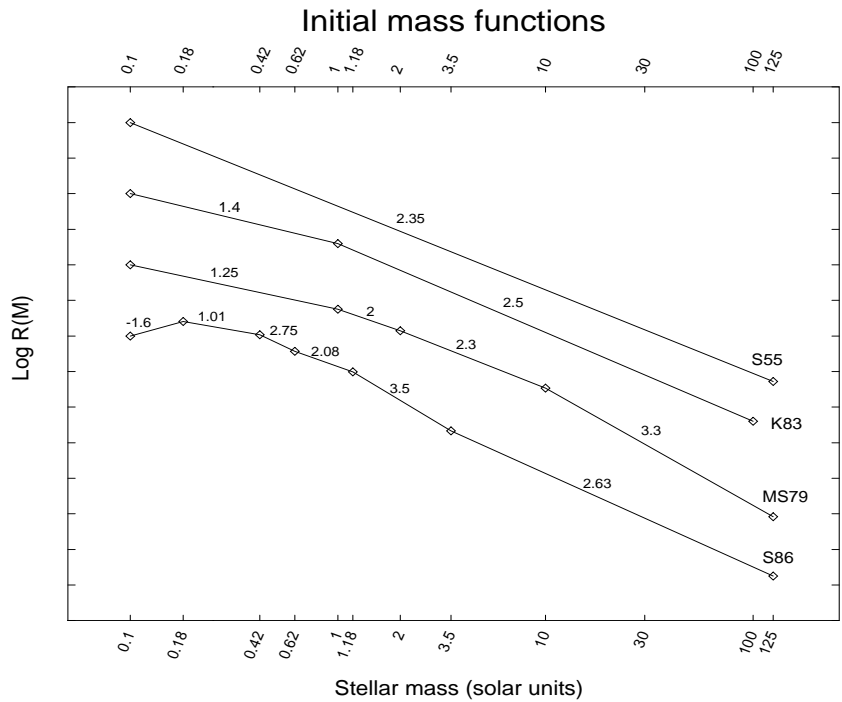}
\caption{\protect \footnotesize{Different IMFs by several authors.
The power law coefficients are indicated close to each 
segment of the relations.
The tick marks on the abscissa are one unit apart, i.e., each tick mark 
corresponds to a factor of 10 in the number of stars formed per unit mass.
The designations are as follows: S55 $\equiv$ Salpeter (1955), K83 $\equiv$ 
Kennicutt (1983),
 MS97 $\equiv$ Miller \& Scalo (1979), S86 $\equiv$ Scalo (1986). 
The MS79 and S86 IMFs
are shown as parameterized by Bruzual \& Charlot (1993).}}
\label{fig_IMF}
\end{figure}

 The star formation process is characterized by
two main parameters: the initial mass function (IMF) and the
total star formation rate (SFR).
 First introduced by Salpeter (1955), the IMF
 is usually described as a power law in the range 
of 2--2.5 (the original value proposed by Salpeter is 2.35). Other
characteristics of the IMF are the low and high mass limits,  
usually taken as 0.1M$_\odot$ and 60--120M$_\odot$. 
Several typical IMFs describe star formation in 
different types of galaxies and environments, such as the solar neighborhood, 
the Magellanic clouds, etc. Some IMFs (e.g, Miller and Scalo 1979,
and Scalo 1986) were not published originally as power laws, but can
be fitted as power laws with varying coefficients for different mass ranges.
 In Fig.~\ref{fig_IMF}
four different IMFs, used in various models, are shown.

As for the SFR, its time dependence 
 varies dramatically between different galactic types
(e.g., Gallagher, Hunter \& Tutukov 1984; Kennicutt {\it et al.} 1994). 
In early
type galaxies, the star formation usually decays smoothly with time.  
In late-type galaxies, on the other hand, the star formation is
generally more intense, and the SFR is subject to significant
changes on short timescales.

Despite extensive progress in understanding the star formation process,
there are several issues still not fully understood.
In paper I we introduced two open questions
concerning the star formation processes in galaxies:
(1) What are the mechanisms that govern the star formation process,
and how do they depend on the galactic type and environment?
(2) How do the SFR and IMF depend on various galactic properties, such as
interstellar gas density,
morphology of the interstellar gas, metallicity and
the amount of dust in the interstellar medium.

The simplified assumption is that the SFR depends directly on the density 
of the interstellar gas (Schmidt law, Schmidt 1959). Actually, the
observable quantity is the gas 
surface density.  Testing this parameter against the SFR in normal
galaxies indicates that there is a threshold gas surface density, below
which no star formation takes place (Kennicutt 1989). The value of this 
threshold surface density varies from galaxy to galaxy, thus it is not
a global parameter, though it is believed to be of order a few times
$10^{20} atoms/cm^2$.

Larson (1987) counts four major mechanisms that induce
star formation in galaxies:
large scale gravitational instabilities of gas clouds,
 compression of interstellar gas clouds due to the passage of
 a gravitational density wave,
 compression in a rotating galactic disk - shear forces acting
on the clouds due to differential rotation and
  random collisions between clouds.
In addition, star formation can be triggered by external
influences, such as tidal interaction induced by
another galaxy during a close encounter of two galaxies, or by interaction
with interstellar matter, during a passage 
near a cluster core. 

In order to investigate the above questions, we have constructed four samples
of galaxies, described in paper I.
These galaxies are all dwarfs in the Virgo cluster. The aim in selecting dwarf
galaxies is to eliminate some of the mechanisms described above, i.e. the
grand design density wave and shear differential rotation forces do not act
in dwarf galaxies. Therefore the theoretical situation is simplified.

Our goal in this paper is to derive the star formation properties of the
sample galaxies from the observational data gathered in paper I,
 and to check these againts various star formation scenarios. This will enable
to test the feasibility of several star formation mechanisms.

Since all methods used to determine galactic star formation properties
combine observational data with theoretical models, the
result is model dependent. Realistic models which use conventional IMFs and
reliable stellar evolutionary tracks 
would result in SFR values that may differ usually by at most $\sim$50\% from
each other, given the same observational data. This is approximately the
accuracy of our observed parameters from which the SFR is derived.
As will be discussed below, the uncertainty
is mainly due to effects of dust extinction
and other observational biases.

In order to estimate the SFR in a sample of galaxies, one needs to know the IMF. 
This can be found by fitting a number of observed properties to 
a set of models of different stellar populations, with different IMFs
 (population synthesis).  Since the colors of a stellar population
depend also on its age and history, a specific star formation
history is usually assumed, and the observed properties can be tested against
different IMFs. Naturally,  this approach can be used only for entire samples
 of galaxies and not for individual galaxies. 

Once the IMF is determined (or assumed), the SFR can be calculated on
the basis of models + observations. All the observational techniques 
basically measure the number of massive stars in the galaxy
to trace the 'current' star formation. 
If the lifetime of stars with a certain mass $M$ is 
$T(\!M\!)$
and the star formation rate at this mass is $R(\!M\!)$, then the number of
such stars currently observed is simply 
$N(\!M\!)=R(\!M\!)\times T(\!M\!)$, where both
$R(\!M\!)$ and $N(\!M\!)$ are given per unit stellar mass.

The total SFR is calculated by  
 extrapolating from the massive stars to the entire mass range, using
the IMF. The higher the mass of the stars we use,
the larger the error of the total SFR that can be introduced due to the extrapolation.
On the other hand, concentrating on
 higher mass stars leads to a more 'up to date' star formation result. 
Therefore, there is a tradeoff between 
how `current' the derived SFR is, and the accuracy of the total SFR.
 In paper I we described the
observations of UV and H$\alpha$ line radiation from the sample galaxies,
which we use here for determination of their SFR.

The relation between the number of observed LyC photons and the SFR can
be reduced to the knowledge of the dependence of
$R(\!M\!)$ on $M$, and the number of LyC photons emitted by a 
star with mass $M$
during its entire lifetime $P(\!M\!)$.
With these two parameters, the number of photons emitted
per unit time by an $M$ mass star (per unit stellar mass) is 
$F(\!M\!)=P(\!M\!)\times R(\!M\!)$, and the
total flux of LyC photons is given by:

\begin{equation} \label{e_Nly}
N_c = \int_{M_{min}}^{M_{max}} F(\!M\!)\; dM = 
\int_{M_{min}}^{M_{max}} P(\!M\!)\cdot R(\!M\!) \;dM
\end{equation}

 where $M_{min}$ and $M_{max}$ are the low and high mass limits of the IMF, 
 and the dependence of $R(\!M\!)$ on $M$ is the IMF. 

We can now express the number of LyC photons as:
\begin{equation} \label{e_NsR}
N_c = S\!F\!R \; \frac{{\displaystyle \int_{M_{min}}^{M_{max}} P(\!M\!) \cdot
I\!M\!F(\!M\!)\; dM}} 
{{\displaystyle \int_{M_{min}}^{M_{max}} M \cdot I\!M\!F(\!M\!)\; dM}}
\end{equation}

where $S\!F\!R$ is in M$_\odot$/yr, and the integrals in the equation
originate from the theoretical 
models. We therefore obtain a direct relation between the SFR and the number
 of LyC photons.

This treatment assumes a constant IMF. However,
 some studies indicate a dependence of the IMF on the 
metallicity of the galaxy (e.g., Terlevich \& Melnick 1983). The 
metallicity of a galaxy gradually increase wirh time,
thus, the IMF will also depend on the galactic age.
 In this case the SFR and IMF would be coupled and will be changing
in time.

The LyC flux is derived here through its influence on the ambient hydrogen, e.g.,
by measurement of the resulting Balmer lines and assuming case B recombination. 
For the calculation of the SFR we use the H$\alpha$ line.
 This line has been used by many (Kennicutt 1983,
hereafter K83,
Kennicutt \& Kent 1983; Gallagher, Hunter \& Tutukov 1984; Pogge and
Eskridge 1987; Kennicutt {\it et al.} 1994), mainly because of its high
intensity. A few percent of the total ionizing flux are reemitted as H$\alpha$
(Kennicutt 1989), so it is convenient for tracing the star formation in
faint galaxies. The internal dust extinction influences the line intensity
significantly, and will be discussed in more detail in the following section.

    \section{Observational data} \label{sec_OBS}

In paper I we described the observations of our sample of late-type
dwarf galaxies in the Virgo cluster. Briefly, these are 
broad-band imaging in B, V, R and I, and in rest frame H$\alpha$.
 We explained there how the data
was calibrated and presented lists of magnitude and colors for
the galaxies.
In addition, we observed some of the galaxies in UV and collected FIR
data from IRAS. 
%
%
%
%

The observational data presented in paper I were not corrected for 
internal dust extinction.
Such a correction is possible
with the use of far infrared (FIR) data, radio continuum 
data, or full spectral 
information for the object in question. None of these is available for
the sample galaxies, except for IRAS data which, in most cases, consists
only of an upper limit. Moreover, in this type of galaxies the FIR 
radiation from dust is related
more to the LyC flux in the H II regions than to the amount of dust, 
thus any correction relying on the IRAS data would be unreliable.
We thus prefer to
estimate a typical dust extinction for the entire sample, rather than
correct for each object individually.

Calzetti {\it et al.} (1994) have investigated the dust extinction in
a sample of starburst and blue compact galaxies. They characterize the
amount of dust by the difference of optical depths of the Balmer H$\alpha$
and H$\beta$ lines ($\tau_b^l$). This is derived from the difference between
the observed and the theoretical values of the line ratios, where they assumed
a case B Balmer decrement. They find that $\tau_b^l$
spans the range from $\sim$zero to $\sim0.9$. Though the range is wide, 
we believe it
is useful for statistical purposes to adopt a typical value. We adopt
$\tau_b^l=0.4$ as a typical dust estimator in our sample galaxies. This
gives a continuum H$\beta$ -- H$\alpha$ `color excess' of 
E(H$\beta$ -- H$\alpha$) = 0.21$\;mag$.
The extinction law derived by Calzetti {\it et al.} (1994) is not significantly
different from the standard Savage \& Mathis (1979, SM) law in the optical
regime. Using it and the value of $\tau_b^l$, or the corresponding
E(H$\beta$ -- H$\alpha$), we obtain the typical
correction for the various colors used here:\\

    \begin{tabular}[b]{|c|c|c|c|c|}
\hline
    Color & B--V & V--R & R--I & UV--V $^*$ \\
\hline
    Color excess  & 0.17  &  0.1 & 0.13 & 0.92 \\
\hline
    \end{tabular}
\\
 $*$ -- "UV" implies here 1650\AA, as will be explained later

\vspace{5mm}

The effect of dust extinction on the H$\alpha$ line results
 is subject to larger errors than that of the galactic colors.
 This is because we need total extinction to correct the
observed line flux, which is larger than color excess, and
so is its error. 
For a general estimate we use the standard value of the 
total-to-selective extinction
$\frac{A_V}{E(B-V)}=3.1$ and obtain for E(B--V)=0.17: $A_{H\alpha}=0.42$
and $A_V=0.53\;mag$. As will be explained in the following section, 
$A_{H\alpha}=0.42$ does not represent the actual extinction of the 
H$\alpha$ line, but rather that of continuum photons
near 6563\AA.

%

Despite our small sample, it is important to consider the relation between
 UV radiation and H$\alpha$ line strength of the galaxies. This is
because the UV radiation shortward of 2000\AA\ is dominated 
by young OB stars associated with HII regions.
 The H$\alpha$ radiation is correlated with the
UV radiation shorter than 912\AA, while our UV data is at longer wavelengths,
but still describing the young stellar population.
For this, [UV--55] monochromatic colors were calculated, where
UV is represented by the FAUST $\sim$1650\AA\ band, and 55 represents the
monochromatic magnitude at 5500\AA. For
galaxies with IUE data, the two IUE results from paper I were interpolated to
derive approximate monochromatic magnitudes at 1650\AA\ (the region 1500--1700\AA\
 is very noisy
in some IUE spectra). The error was taken
as the larger of the two IUE errors. 

\begin{figure}[htbp]
\vspace{7.5cm}
\includegraphics{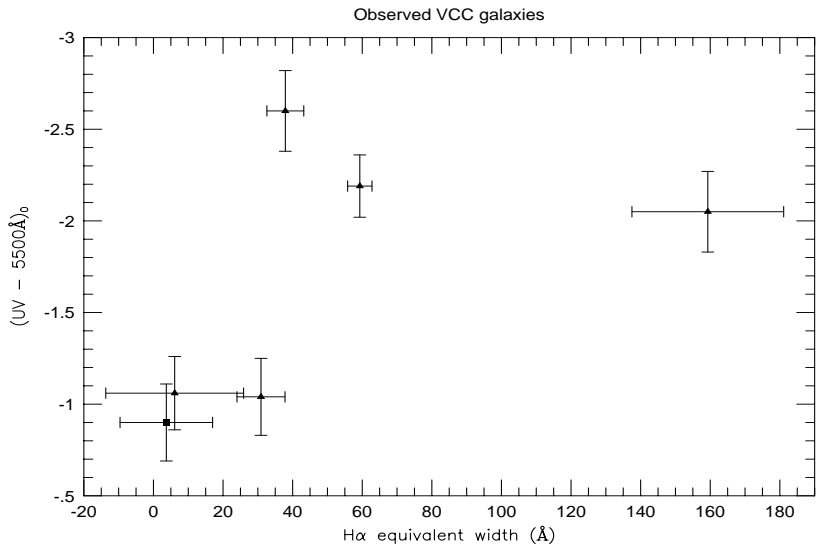}
\caption{\protect \footnotesize{[UV--55] monochromatic colors 
vs. the H$\alpha$ equivalent width of the sample galaxies.}}
\label{fig_UVH}
\end{figure}

The [UV--55] values were corrected for internal 
extinction as described above.
Fig.~\ref{fig_UVH} displays this [UV--55] color versus the H$\alpha$
equivalent width of the galaxies. The two points in the upper left of the
figure are FAUST points, whereas the others are IUE points.  
The apparent lack of correlation may be due to different geometries of
the dust and HII regions within the galaxies. 

%
%
%
%
 
     \section{Discussion} \label{sec_dis}

\subsection{Star formation rates} \label{sub_SFR}


  For the calculation of the SFR we adopt the model of Gallagher, Hunter \& 
Tutukov (1984, GHT).
We use it mostly because it was tested both on spiral galaxies
 and irregular galaxies, which
are more similar to the VCC galaxies. GHT use a Salpeter (1955) IMF: 
$N(m)\propto m^{-2.35}$
with a lower mass cutoff of 0.1 M$_{\odot}$, and an upper mass cutoff of 
100 M$_{\odot}$. This gives the relation between
the total SFR and the flux of LyC photons, as described in 
equation~\ref{e_NsR}. 
 GHT adopt a standard
gas temperature of 10000K and an efficiency of 2/3 of reemission of the
ionizing stellar flux by the HII region, i.e., one third of the ionizing
photons are lost to dust before ionizing the hydrogen or escapes from the
HII region. The total SFR is:

\begin{equation} \label{e_HaSFR}
 SFR = 1.27\times10^9 F(H\alpha) D^2
\end{equation}
where the SFR is given in M$_{\odot}$/yr, $F(H\alpha)$ is the reddenning-free
line flux in $erg/cm^2/s$, and D is the distance to the object in Mpc. 
The H$\alpha$ flux depends very weakly on the gas temperature. It
changes by 4\% in the range of 5000 -- 20000K and is practically  
independent of the electron density.
One should bear in mind that other models with different IMFs, stellar 
evolutionary tracks, etc., yield 
values that may differ by a factor of two from this one 
(Kennicutt {\it et al.} 1994). 

The most important effect that reduces the
observed H$\alpha$ flux is extinction by dust, both of the ionizing photons
themselves and of the H$\alpha$ photons. In addition, the ionizing photons
may escape the HII region if the covering factor of the clouds is less than
unity or the optical depth of the clouds is low. 
This will further reduce the resultant H$\alpha$ flux. However,
a common assumption is that only
a small fraction of the ionizing photons is extinguished by the dust or
escapes from the HII region
(GHT use 1/3 as mentioned, Mas-Hesse \& Kunth (1991) also use 1/3, but
consider a value of up to 3/4 for the dust extinction). This is because 
the covering factor is usually believed to be close to unity and the HII
region $\sim$ionization-bound (e.g., Shields 1990),
 and within it the hydrogen absorbs very efficiently the ionizing photons. 
Another reason may be the location of
the dust relative to the HII region. Calzetti {\it et al.} (1994) find that the
picture of dust being mixed with the young stars inside the HII region
can almost certainly be ruled out, and the dust probably lies outside the
line emitting region, forming a shell around the newly formed
stars. This means that the dust expresses itself mostly by removing the
H$\alpha$ photons from our line of sight, and affecting less the LyC radiation.

In the previous section we obtained a typical value for the extinction at 
the H$\alpha$ wavelength of $A_{H\alpha}(c)=0.42\;mag$. This is for the {\it
 continuum} photons at this wavelength, and is smaller than the actual 
extinction of the line itself. The reason is that the continuum and 
line photons
come from different areas in the galaxy, and the dust, residing mostly around
the HII regions, affects less the light from other parts of the galaxy. 
Calzetti {\it et al.} (1994) quote $\tau_B^c = (0.5 \pm 0.11)\tau_B^l $, 
where $\tau_B^c $ is the difference between the optical depth for 
{\it continuum} photons at the wavelength of H$\beta$ and H$\alpha$, 
and $\tau_B^l $ is this difference for {\it line} photons.
 This means that the optical depth of the dust towards an HII
region is twice its value towards other parts of the galaxy. Our
correction for extinction should therefore be of 0.84$\;mag$. In order not to 
overestimate the SFR we choose a somewhat smaller value - we assume that the 
dust reduces the observed H$\alpha$ flux by a factor of two, which corresponds 
to $A_{H\alpha}(l)=0.75\;mag$. Equation~\ref{e_HaSFR} then becomes:
\begin{equation} \label{e_HaSF2}
 S\!F\!R = 2.54\times10^9 F(H\alpha) D^2
\end{equation}
where now $F(H\alpha)$ is the actual observed line flux of our sample objects.

The common 
distance adopted for all galaxies, in order to derive their total SFR,
was 18 Mpc (see paper 1).
We calculated the total SFR of each of the sample galaxies using
equation~\ref{e_HaSF2}, and the results are presented in Table~\ref{tab_Ha2}, 
  subject to the uncertainty in the distance.
 It is worthwhile,
therefore, to consider a distance-free parameter such as the star formation
per unit galactic area. This can be derived using the size of each galaxy from
its CCD image.
 Following equation~\ref{e_HaSF2}, the ongoing SFR per unit area is given by:
\begin{equation} \label{e_HaSFB}
 S\!F\!R/area = 1.078\times10^8 F(H\alpha)/\Box"
\end{equation}
where the SFR/area is in M$_{\odot}$/yr/$pc^2$ and 
$F(H\alpha)/\Box"$
 is the observed H$\alpha$ flux per square arcsecond, 
as derived in 
paper 1. This parameter is
free of the uncertainty of the distance to the objects, since both quantities
 scale as $D^2$.
 These values are also given in Table~\ref{tab_Ha2}.


\begin{table}[htbp]
\vspace{8.2cm}
\includegraphics{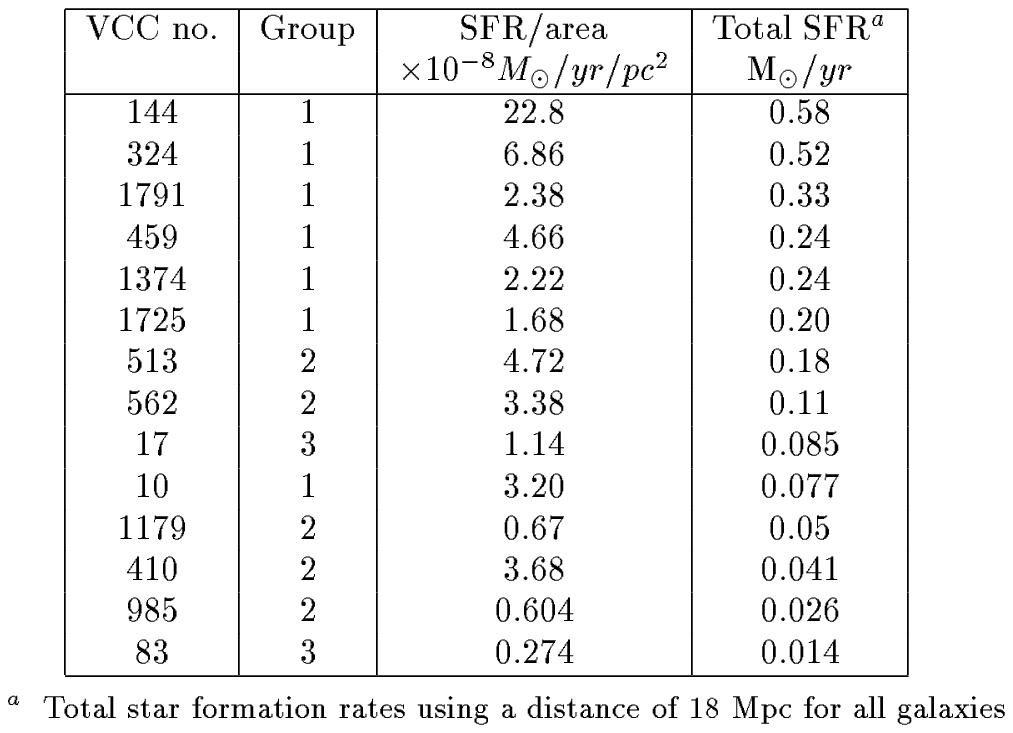}
\caption{\protect \footnotesize{Star formation properties
of VCC galaxies in decreasing order of H$\alpha$ line flux.}} 
\label{tab_Ha2}

\end{table}


Not surprisingly, the galaxies undergo strong star formation per unit
galactic area, typically of a few times $10^{-8} M_\odot/yr/pc^2$, but have
only modest total SFRs, \linebreak of order 0.03 -- 0.3 $M_\odot/yr$. This
emphasizes the small size of these dwarfs. For comparison, our galaxy has a 
total SFR of $\sim 5 M_{\odot}/yr$ but SFR/area of only \linebreak 
\mbox{$\sim 7\times 10^{-9}
 M_{\odot}/yr/pc^2$.} Large, active, star-forming galaxies have SFR/area
similar to the sample galaxies (e.g., Pogge \& Eskridge 1987). 

This interpretation implies that mechanisms 
believed to be responsible for star
formation in large galaxies but not in dwarfs, such as compression by 
density waves or rotational shear forces, are not much more efficient
than other mechanisms, such as random collision between clouds or gravitational
instabilities caused by other effects. Otherwise, there would be significantly
 more
star formation/area in large star-forming galaxies than in late-type dwarfs.
It is worth mentioning that here the {\it efficiency} of the star formation
process, caused by the various mechanisms, is tested, rather than how wide
spread 
these mechanisms are. This is because our sample of galaxies was selected
{\it a-priori}
 as star-forming galaxies. In order to investigate the
occurrence of the mechanisms, namely how frequently star formation is caused
by any specific  mechanism, all galactic types would have to be investigated 
simultaneously, a procedure which is beyond the scope of this work. Therefore, 
the conclusion at this stage may be
 that once star formation occurs, its efficiency is approximately 
the same, regardless of the mechanism which induced it.

\begin{figure}[htbp]
\vspace{7.5cm}
\includegraphics{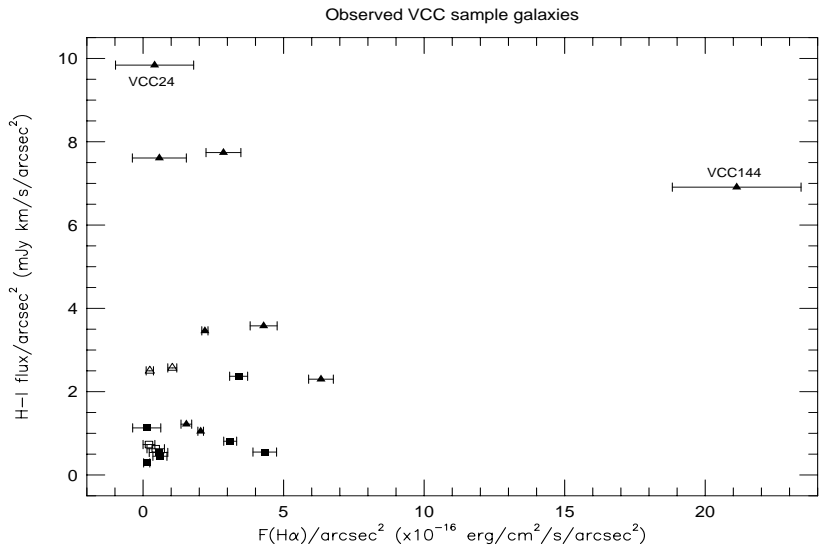}
\caption{\protect \footnotesize{HI vs. H$\alpha$ surface brightnesses
of the sample galaxies.}}
\label{fig_HIHa}
\end{figure}

As for the hydrogen content, there is no apparent dependence of the SFR/area 
on the subsample to which the galaxy belongs,
but the high HI subsample has larger {\it total} SFR. This
implies that our selection of high or low HI subsample naturally selected 
larger or smaller galaxies on average, and there is no strong correlation
between the neutral hydrogen and the SFR. 
It is worthwhile, therefore, to test the HI surface
brightness rather than the total HI flux. This relation is displayed in 
Fig.~\ref{fig_HIHa}, and no significant correlation is visible
between the HI and H$\alpha$ surface brightnesses.

This finding is
in agreement with the SFR being dependent on the {\it total}
amount of hydrogen (atomic + molecular) and almost independent on
each of these alone (e.g., Buat {\it et al.} 1989). The SFR depends on many 
parameters, therefore the influence of the neutral hydrogen component cannot
be clearly isolated.

\subsection{IMF and star formation histories} \label{sub_SFH}

The broad-band colors of galaxies are not very sensitive to their star 
formation properties, such as the IMF and star formation history, thus it is
useful to consider other galactic properties. 
As pointed out by K83, a combination of broad-band data with H$\alpha$ can
provide information about the IMF of the objects. 
Kennicutt {\it et al.} (1994) treat
a set of models, with updated stellar evolutionary
 tracks, for star-forming galaxies,
which derive broad-band colors and
H$\alpha$ equivalent widths for various IMFs and star formation histories.
In order to compare our data with these models we compensate for 
the dust extinction. The measured broad-band colors were corrected as described above.
The influence of dust on the measured
H$\alpha$ equivalent width is more complicated because it depends on the geometry of
the HII regions in the galaxy. As discussed above, the optical depth of the 
dust towards the HII regions is believed to be twice as large, on average,
 than the value towards continuum
emitting regions. Since the measured EW of a spectral line is the ratio of 
line-to-continuum radiation, the EW will be 
reduced by the same factor as the continuum radiation at its wavelength.

\begin{figure}[htbp]
\vspace{14.8cm}
 \includegraphics{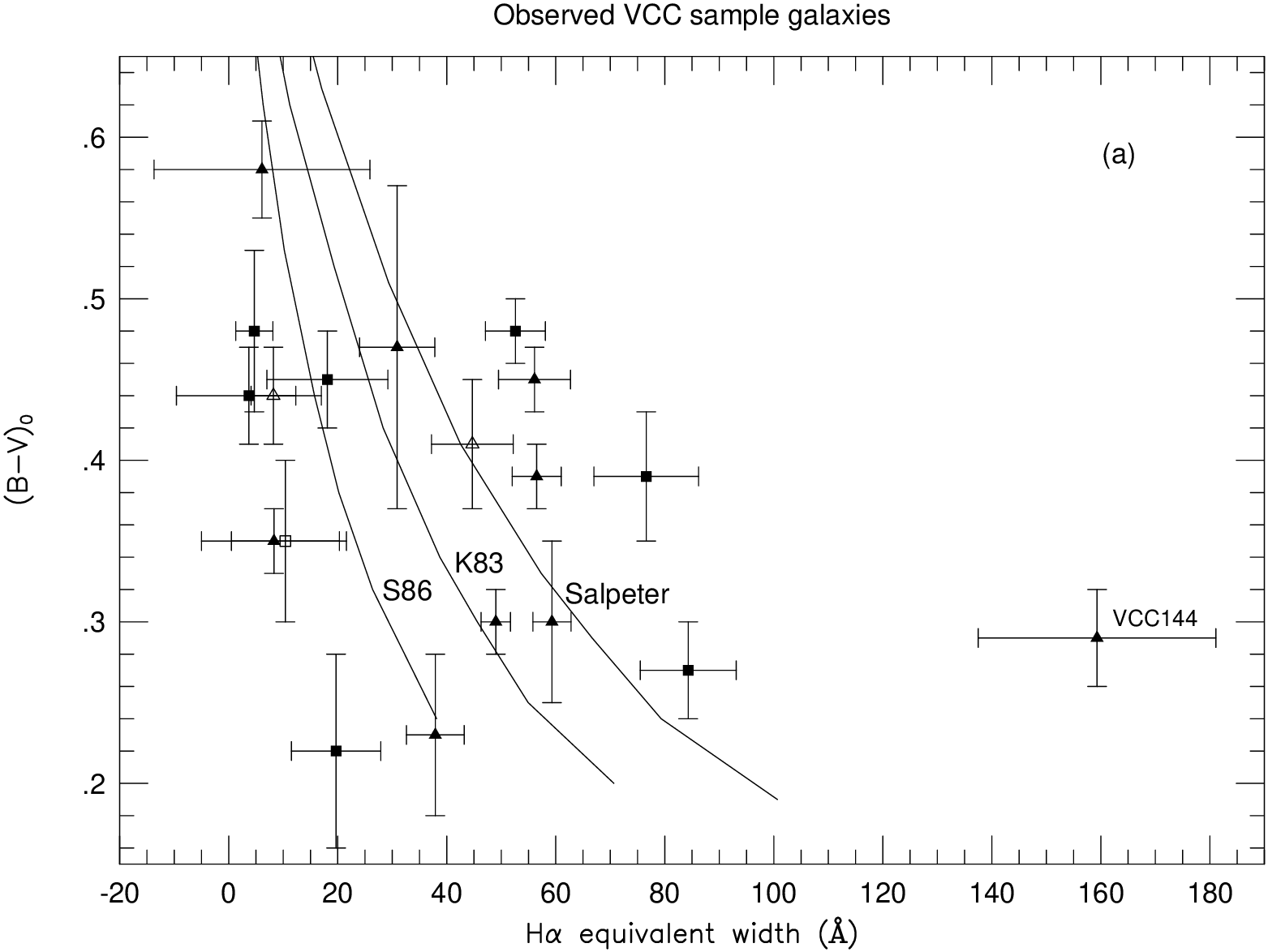}
\includegraphics{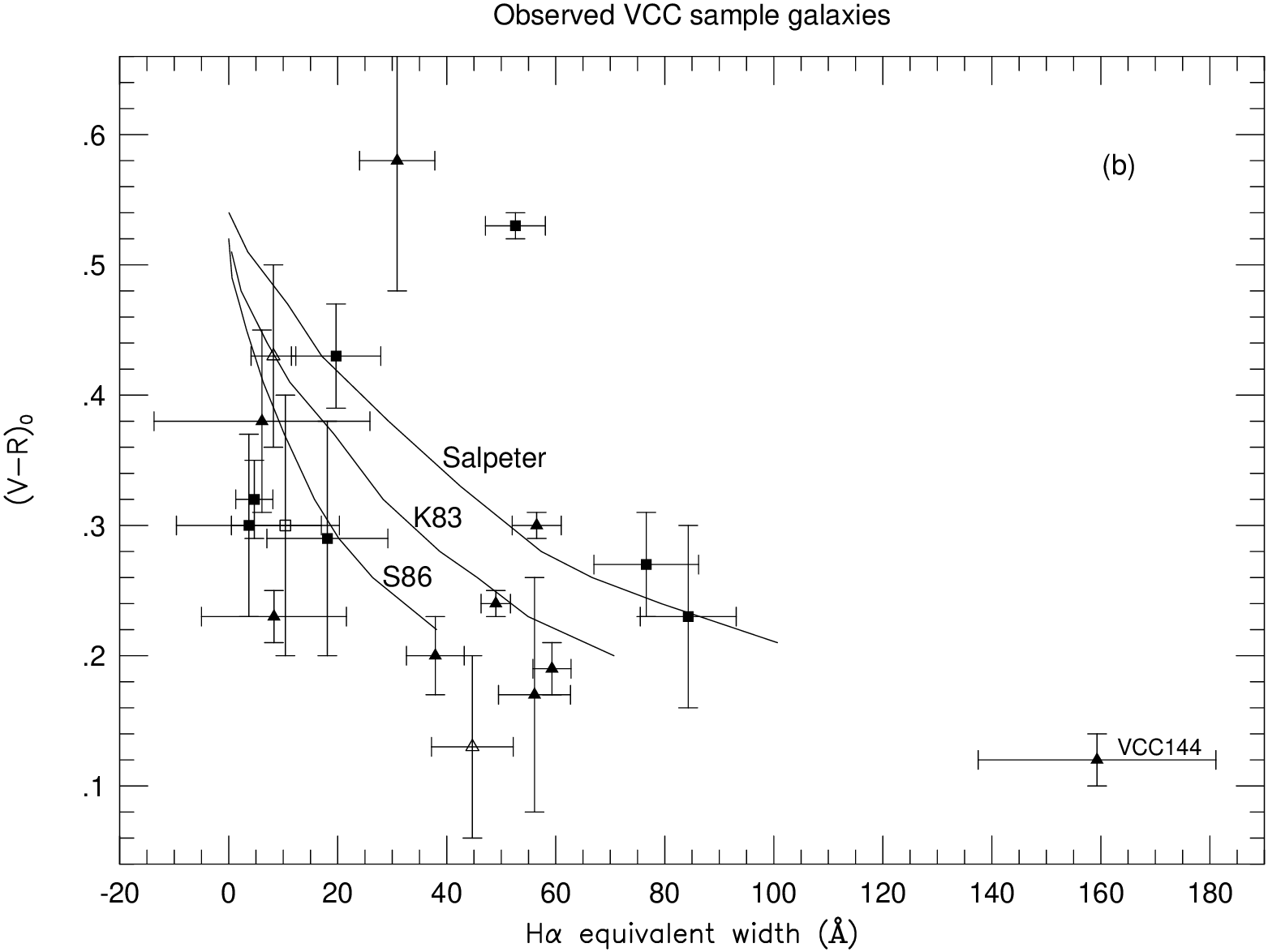}
\caption{\protect \footnotesize{Broad-band colors vs. H$\alpha$ equivalent 
widths for the sample
galaxies, together with models from Kennicutt {\it et al.} (1994). B--V
 is displayed in (a)
and V--R in (b). The three models have lower and upper stellar mass limits
of 0.1M$_\odot$ and 100M$_\odot$, and are marked as follows: Salpeter: 
$\alpha$ = -2.35, K83: $\alpha$ = -1.4 below 1M$_\odot$ and -2.5 above 
1M$_\odot$, S86: tabulated values from Scalo (1986).}}
\label{fig_BVH}
\end{figure}

For a typical extinction at H$\alpha$ continuum wavelength of 
0.42$\;mag$,
the EW[H$\alpha$] is reduced by a factor $\sim$1.5. We adopt this value
for comparison with the models. Since we would still like to display observed
values of equivalent widths, we prefer to scale the values from
Kennicutt {\it et al.} (1994) by this factor, rather than change the 
observed data.
The EW[H$\alpha$] displayed in Figs.~\ref{fig_BVH} are, thus, 
observed values, while the models are scaled according to the factor described here.
The broad-band colors are the generally corrected ones, compared with the
original model values of Kennicutt {\it et al.} (1994).

Figs.~\ref{fig_BVH}a and~\ref{fig_BVH}b show three IMFs
which differ in their \mbox{slope $\alpha$.} 
Within each model, the galaxy has a fixed age of 10 Gyr with various
star formation histories, which are expressed
by different $b$ parameters. The $b$ parameter is the ratio between the current 
SFR and the average past SFR. Although the galaxies
follow a general trend which correlates bluer colors with higher star formation
rates, no specific IMF among the three
can be clearly assigned to the galaxies. This probably
arises from two major reasons: 
\begin{enumerate}
\item The observational points are dispersed due to 
(ordered by importance) - 
different amounts of extinction, different contributions of the [N II] lines, and
different wavelength shifts, relative to the peak-transmittance of the H$\alpha$
filter. 
\item The models have all the same age of 10 Gyr. BCD galaxies may
 be young (e.g., Gondhalekar {\it et al.} 1984) and, in general, our sample 
galaxies should span a range of ages,
therefore the models cannot depict the current situation for the entire sample.
\end{enumerate}
Thus, the IMF of the sample galaxies (assuming that they {\it do} share the 
same typical IMF),
cannot be clearly deduced using these data, and neither can their ages,
although the data are consistent with the models.

\begin{figure}[htbp]
\vspace{14.8cm}
 \includegraphics{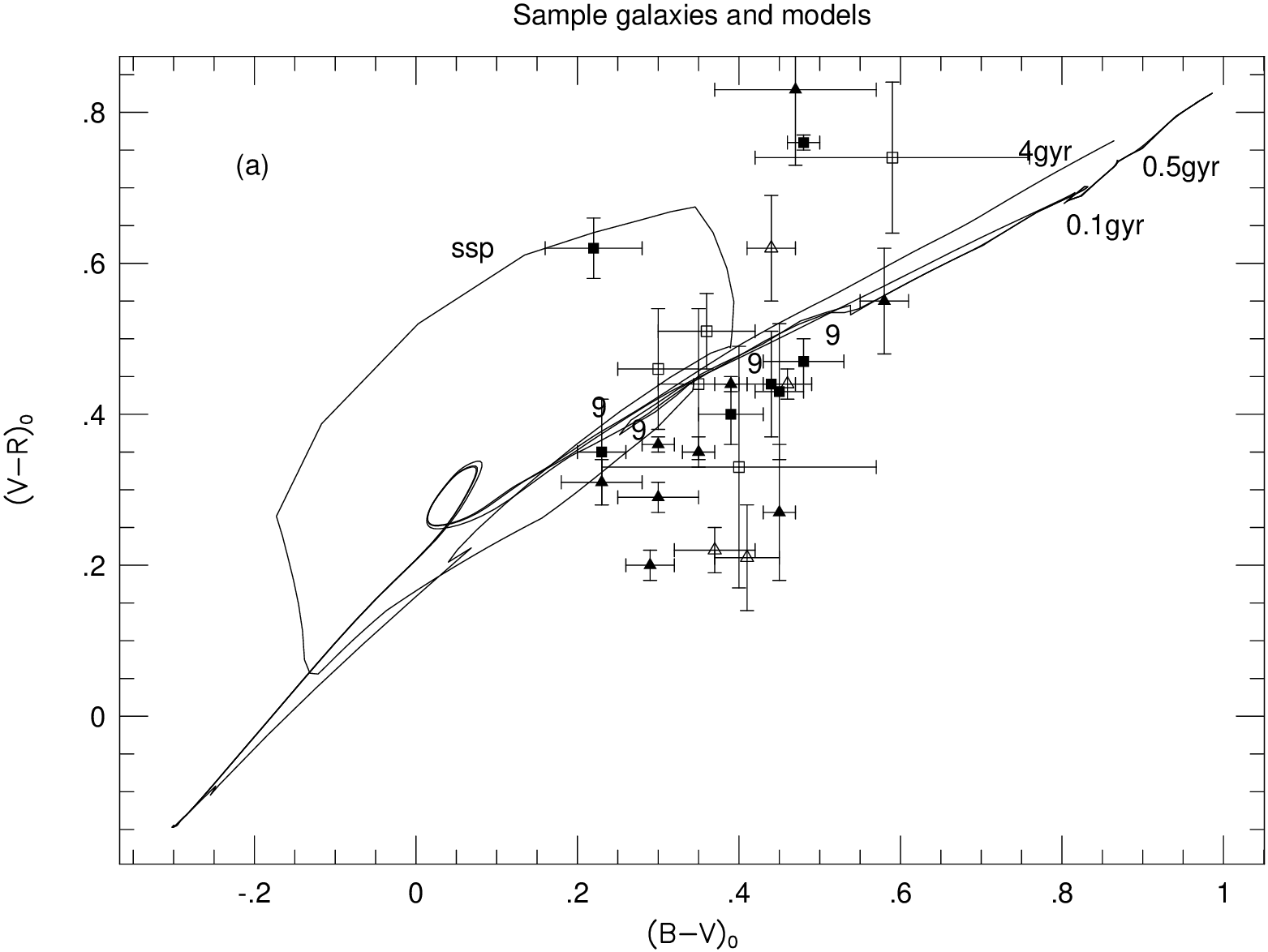}
\includegraphics{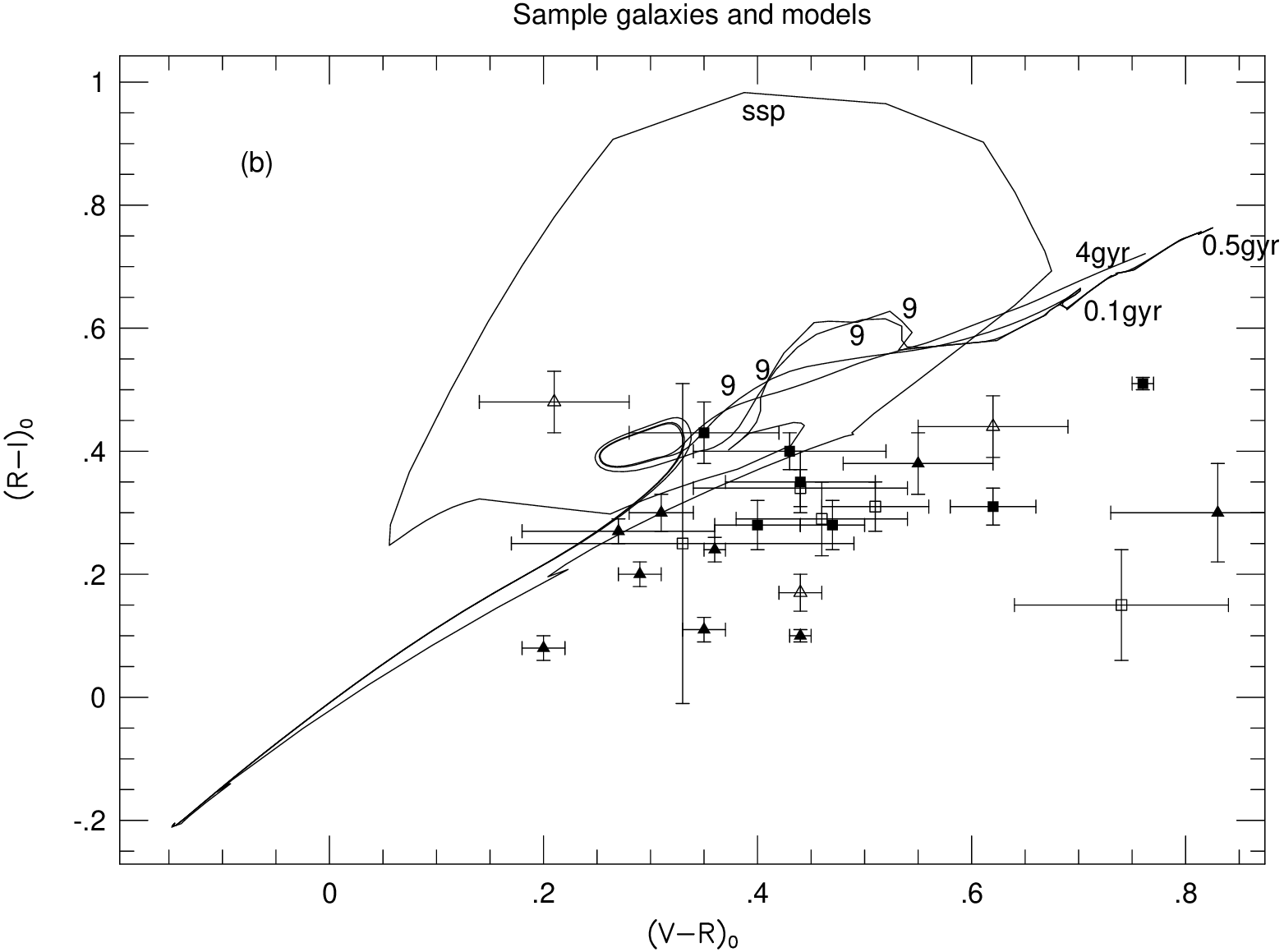}
\caption{\protect \footnotesize{Color-color diagrams of the sample galaxies, 
with various models
from BC93, differing by star formation histories: the 
decay time of each model is
indicated. {\bf ssp} means an instantaneous burst of star formation. 
`9' indicates
the point where the age of the model is 1 Gyr. All the models have a Salpeter
mass function (with a mass range of 0.1--125M$_\odot$). 
Other IMFs do not differ significantly from this and
are not depicted here.}}
\label{fig_BVmod}
\end{figure}

It is interesting to compare the colors of the galaxies with detailed 
population synthesis models, such as by Bruzual \& Charlot (1993, BC93). 
These rely on detailed stellar evolutionary tracks and
provide the time dependence of calibrated broad-band colors and other 
properties of various stellar populations with different IMFs and star 
formation histories. Here we use models with exponentially-decaying SFR
with different decay times, as well as
 an instantaneous burst of star formation. This,
we believe, spans all the practical star formation 
histories. We emphasize that the working assumption is that the galaxies
studied here may be explained by a single stellar population and this
assumption is tested by comparing with the model predictions.
For this, we converted the V--R and R--I color indices of 
the galaxies from the Kron-Cousins system to the
Johnson system, which is the system used by BC93, using
the transformation from Fernie (1983).
%
 Color-color diagrams, together with the
models, are displayed in Figs.~\ref{fig_BVmod}a and~\ref{fig_BVmod}b. 
On each curve, a `9'
indicates where on the plot the age of the models is 1 Gyr. 

It is clear that all the models, except for the instantaneous burst, lie
roughly on the same track. Models with longer decay times have bluer colors
for the same age of the stellar population. The galaxies may have 
 different ages and star formation histories, and they are fitted,
 practically,
by all the models, although the fit is rather poor. This confirms 
again that broad-band colors alone do not contain enough information 
to resolve the star formation
properties of the galaxies. Information in the UV or H$\alpha$ is needed,
since these bands are much more
sensitive to the recent star formation, as the ionizing/short band radiation is 
emitted mostly by stars with short lifetimes.

BC93 provide the number of LyC photons at all stages of development. We can use
this to derive a distance-independent parameter for the sample
 galaxies: 
\begin{equation} \label{e_HamV}
[H\alpha - V] = -2.5 \log[F(H\alpha)] - v \ ,
\end{equation}
where F(H$\alpha$) is in $erg/cm^2/s$ and $v$ is the apparent V
magnitude. This gives the ratio of line to continuum 
radiation, which is essentially the same as $EW[H\alpha]$, except for the
wavelength of the continuum radiation.

\begin{figure}[htbp]
\vspace{7.5cm}
\includegraphics{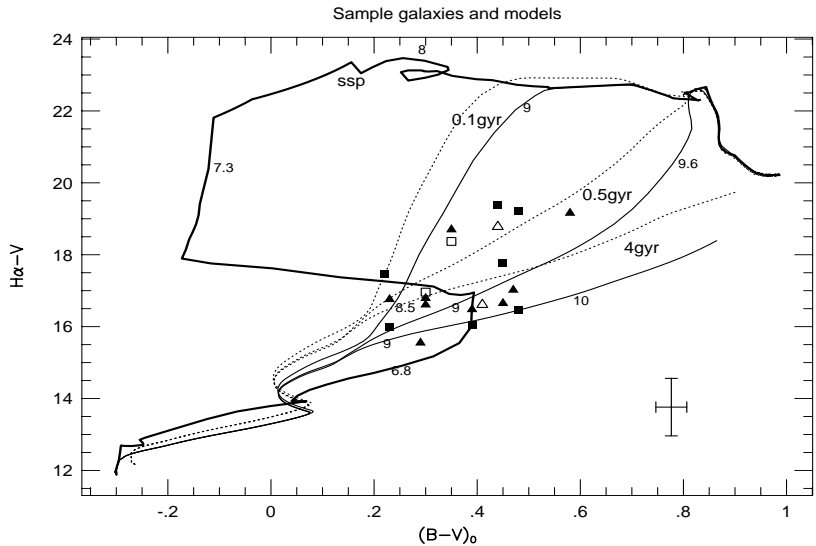}
\caption{\protect \footnotesize{H$\alpha$ --V vs. B--V, with BC93 models.
For each decay time two models are depicted. The solid lines are Salpeter
IMF and the dashed lines represent Scalo's (1986) IMF. The {\bf ssp} model 
with Scalo
IMF is not displayed, as it is very similar to the Salpeter model.
The numbers indicate the log(age) in years only near the Salpeter models, for
clarity. The symbols are the same as in previous color-color plots, 
and a fiducial error bar is at the lower right corner.}}
\label{fig_HBV1}
\end{figure}

\begin{figure}[htbp]
\vspace{7.5cm}
\includegraphics{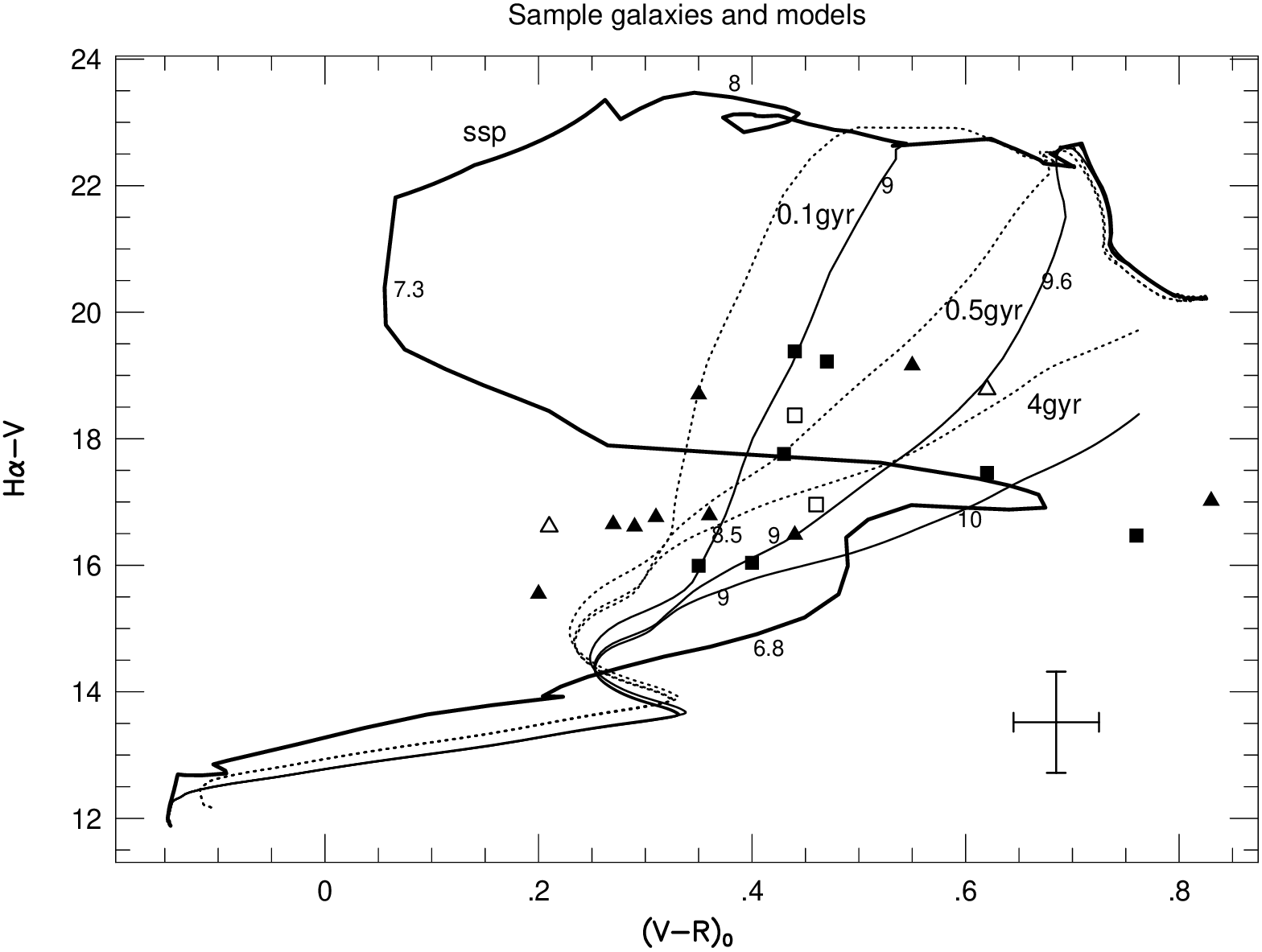}
\caption{\protect \footnotesize{H$\alpha$ --V vs. V--R, with notations the
same as before.}}
\label{fig_HBV2}
\end{figure}

\begin{figure}[htbp]
\vspace{7.5cm}
\includegraphics{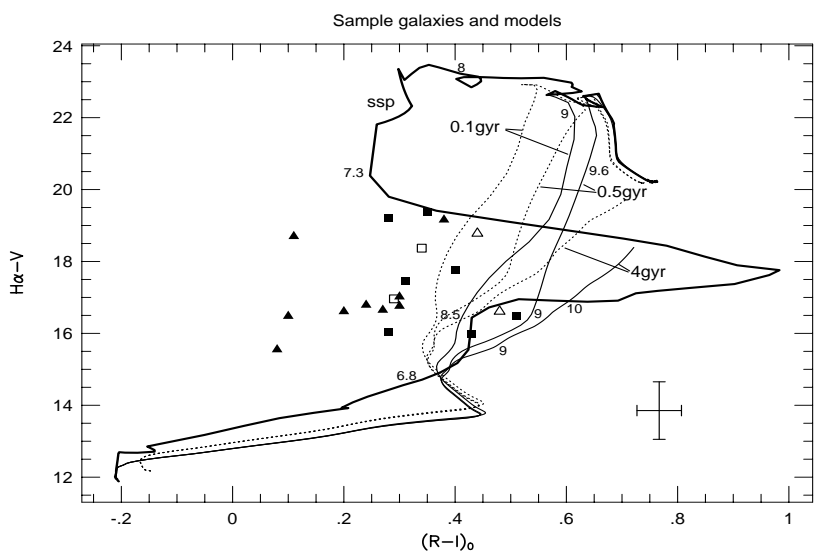}
\caption{\protect \footnotesize{H$\alpha$ --V vs. R--I, with notations the
same as before.}}
\label{fig_HBV3}
\end{figure}

Assuming a simple case B hydrogen recombination theory (Osterbrock 1989), with 
 $T_e\simeq$10000K, the 
H$\alpha$ luminosity is coupled to the flux of LyC photons 
 by:

\begin{equation} \label{e_Nc}
N_c = 7.43 \times 10^{11} L(H\alpha)
\end{equation}
with L(H$\alpha$) in erg/s. Equation~\ref{e_Nc} does not account for
the fraction of LyC photons extinguished by dust, or which escape through 
gaps in the surrounding gas, or through the gas itself, 
as this fraction is relatively small and does not affect our interpretation.
Therefore:

\begin{equation} \label{e_HaV}
(H\alpha-V) = 129.8 - 2.5\; \log[N_c] - V
\end{equation}
where $\log[N_c]$ and V are those provided by BC93. 'Color-color' 
diagrams, with \mbox{(H$\alpha$ --V)} 
as one of the colors, are very useful for tracing the star formation
processes, because the H$\alpha$ luminosity changes by orders 
of magnitudes during
the lifetime of a galaxy. Such diagrams are displayed in 
Figs.~\ref{fig_HBV1},~\ref{fig_HBV2} and~\ref{fig_HBV3}.
 
The figures show
that the B--V colors are consistent with the models, while the V--R and 
R--I colors are bluer than expected. This can also be seen in 
Figs.~\ref{fig_BVmod}a and~\ref{fig_BVmod}b,
 where the points lie, on average, below the 
predicted curves. 
What could the reason for this deviation? A reasonable possibility is
extinction by dust, that reddens the galactic colors.
Extinction effects, however, will move the points more or less along the 
curves of Figs.~\ref{fig_BVmod}, even
when considering different dust geometries and extinction laws, i.e., the
differences which arise due to various dust configurations are insignificant
compared to the deviation seen here. The `color excess' E(H$\alpha$--V), which
was not taken into account in our derivation of (H$\alpha$--V), should
typically be \mbox{$\sim0.3\;mag$,}  according to the considerations
 discussed above.
This is less than the typical error displayed in 
Figs.~\ref{fig_HBV1},~\ref{fig_HBV2} and~\ref{fig_HBV3}, and thus 
does not affect our interpretation. Therefore, dust extinction
is not likely to be the reason for the mismatch between models and the 
observed parameters. 

As for the LyC radiation that may escape from the HII regions due to a 
partial
coverage by hydrogen clouds, or due to their insufficient optical
depth, a realistic assumption would be that at most 50\% of these photons 
can escape. This translates to 0.75 $mag$,
which, together with the 0.3 $mag$ extinction, may reduce by 1.05 $mag$
 the observed value of H$\alpha$--V. This, again, is not
sufficient for bringing the observed results in accordance with the models.

Another effect may be
the influence of a strong H$\alpha$ line on the R band result. This will
make the V--R color redder than the line-free color, and move R--I to bluer
values, an effect that may explain the R--I vs. V--R diagram. In the
V--R vs. \linebreak B--V diagram, however, this effect should move the 
points toward redder V--R color relative to
their original locus, in the opposite direction to what is
seen here. In any case, only the bluer galaxies should show this
effect, which cannot change the colors by more than 0.1$\;mag$, not
enough to explain the deviation seen in Fig.~\ref{fig_HBV3}. In paper I
we have mentioned that our B--V results are redder by $\sim 0.1\;mag$, on 
average, than other published data for some VCC galaxies. This is
not explained by an observational bias, but even if such a bias exists,
 and allowing systematic errors of $0.1\;mag$ or even more in R--I,
we cannot account for the difference between
our results and the various models.
Therefore, we conclude that the galactic broad-band colors and H$\alpha$ 
data of our sample galaxies cannot be explained, as 
a whole, by a single age stellar population model. This rejects, with a significant
degree of confidence, our working assumption.

The spectra of the stellar populations calculated by BC93 are a superposition
of spectra of individual stars, weighted according to the IMF. In young
star-forming regions, however, the contribution of the nebulosity is 
significant, mostly producing the emission lines and also affecting the
continuum. This is not included in models such as
of BC93. Mas-Hesse \& Kunth (1991, MHK) 
calculated observable parameters of starburst regions relying on stellar
evolutionary tracks with Z=Z$_\odot$ and Z=Z$_\odot$/10 metallicities and
using short time steps of $\geq$0.05 Myr.
Their models include,
along with the stellar contribution, the radiation of the 
surrounding gas and dust. They provide, 
among other parameters, H$\beta$ 
luminosities and rough SEDs of 0 -- 20 Myr old burst populations with different
IMFs and metallicities. The contribution of the nebulosity to the continuum
is $\sim$30\% in R and $\sim$10\% in V during the first few $10^6$ years, and is 
more pronounced for low metallicity populations. It is, thus, important to
consider this effect on the broad-band colors.

We calculated from the SEDs given by MHK the
approximate broad-band colors B--V, V--R and R--I, where R and I are
the Johnson bands, by scaling to the
 flux densities at each central wavelength of a normal zero
 magnitude star from Allen (1973). In addition we calculated (H$\alpha$ --V),
using the case B line ratio $H\alpha/H_\beta = 2.86$ to estimate the
H$\alpha$ contribution.

\begin{figure}[htbp]
\vspace{7.5cm}
\includegraphics{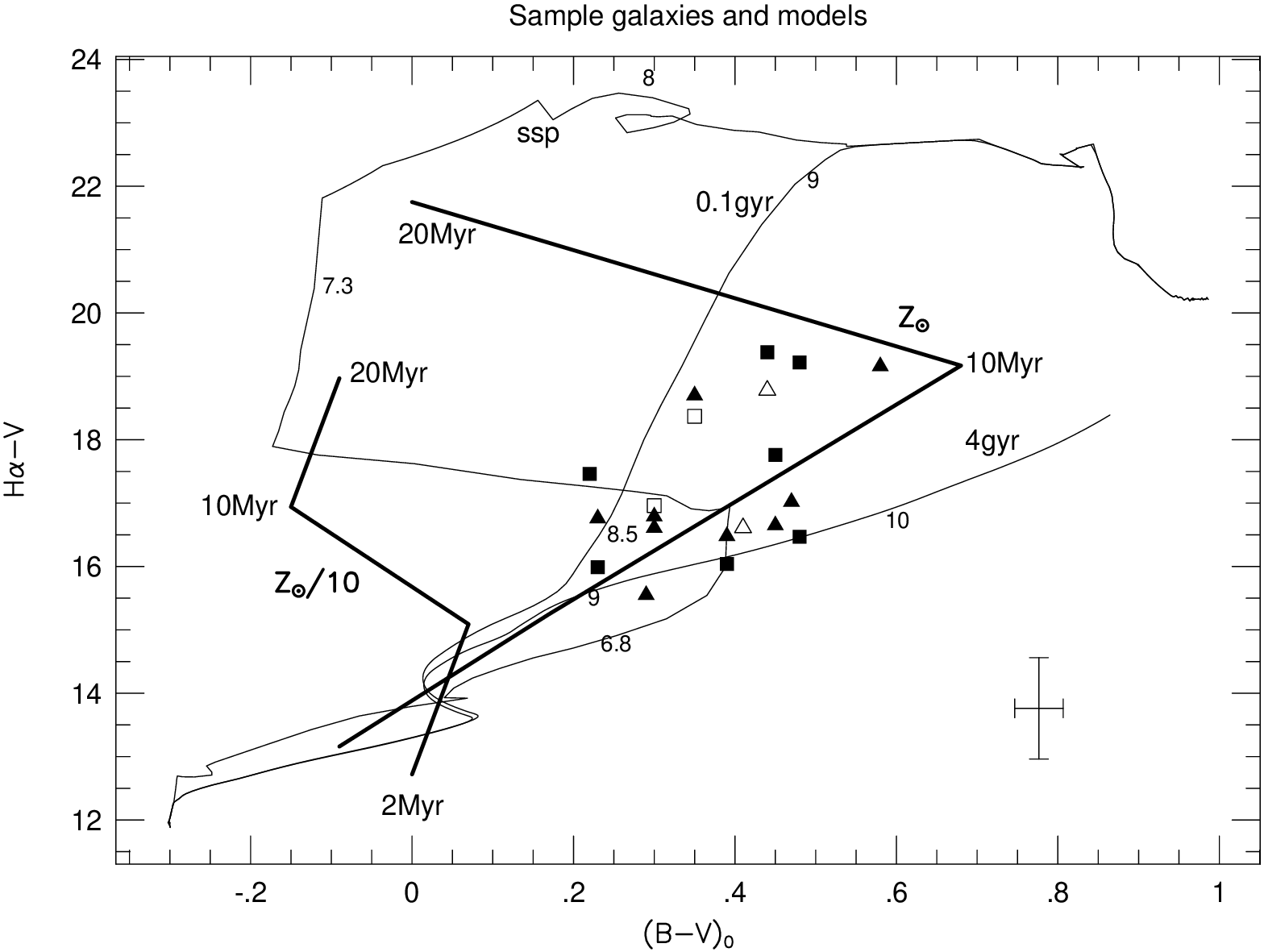}
\caption{\protect \footnotesize{H$\alpha$ --V vs. B--V, with BC93 models and 
Mas-Hesse \& Kunth
(1991) models for solar and low metallicities.}}
\label{fig_HBH1}
\end{figure}

\begin{figure}[htbp]
\vspace{7.5cm}
\includegraphics{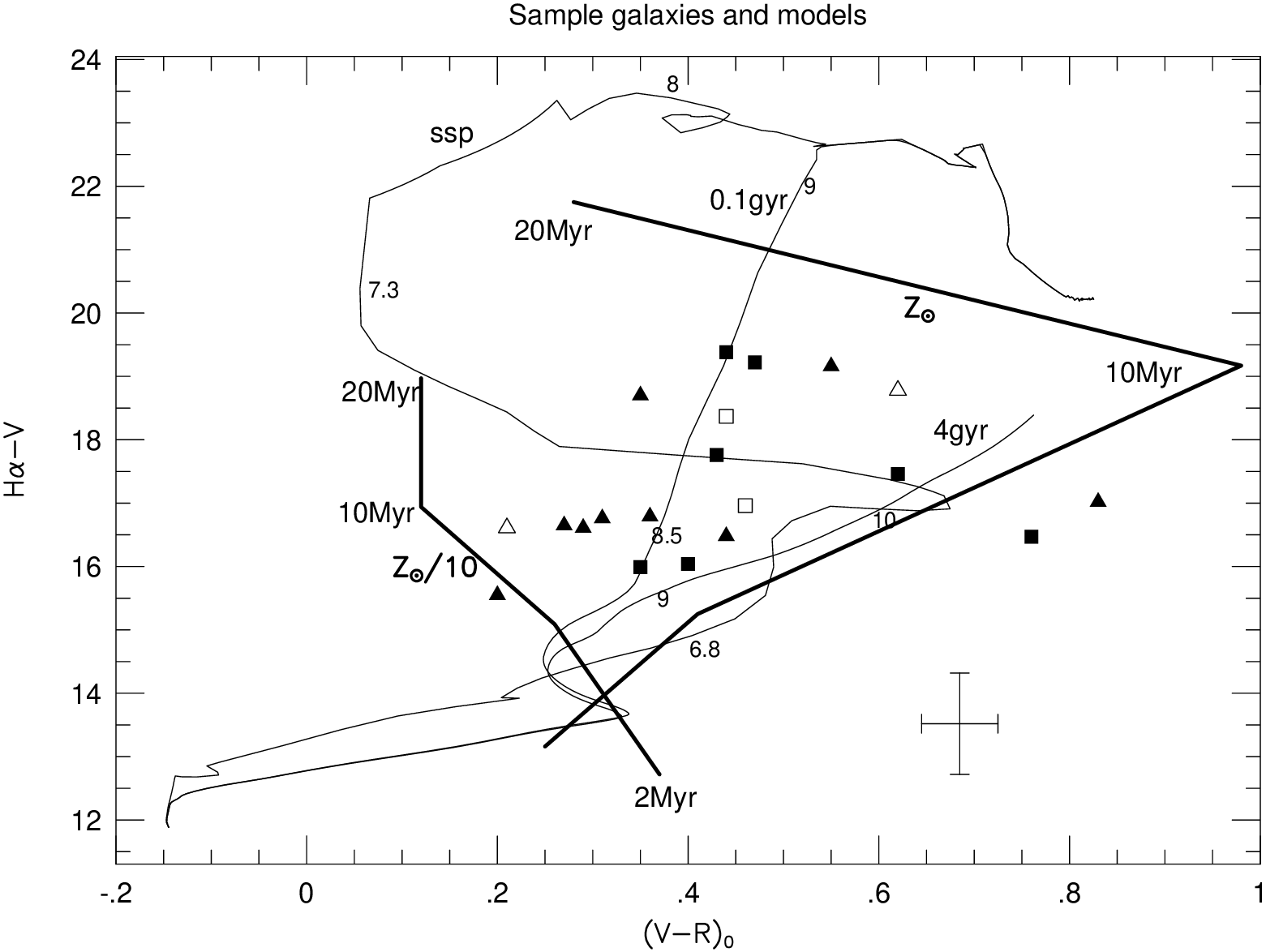}
\caption{\protect \footnotesize{H$\alpha$ --V vs. V--R, with notations the
same as before.}}
\label{fig_HBH2}
\end{figure}

\begin{figure}[htbp]
\vspace{7.5cm}
\includegraphics{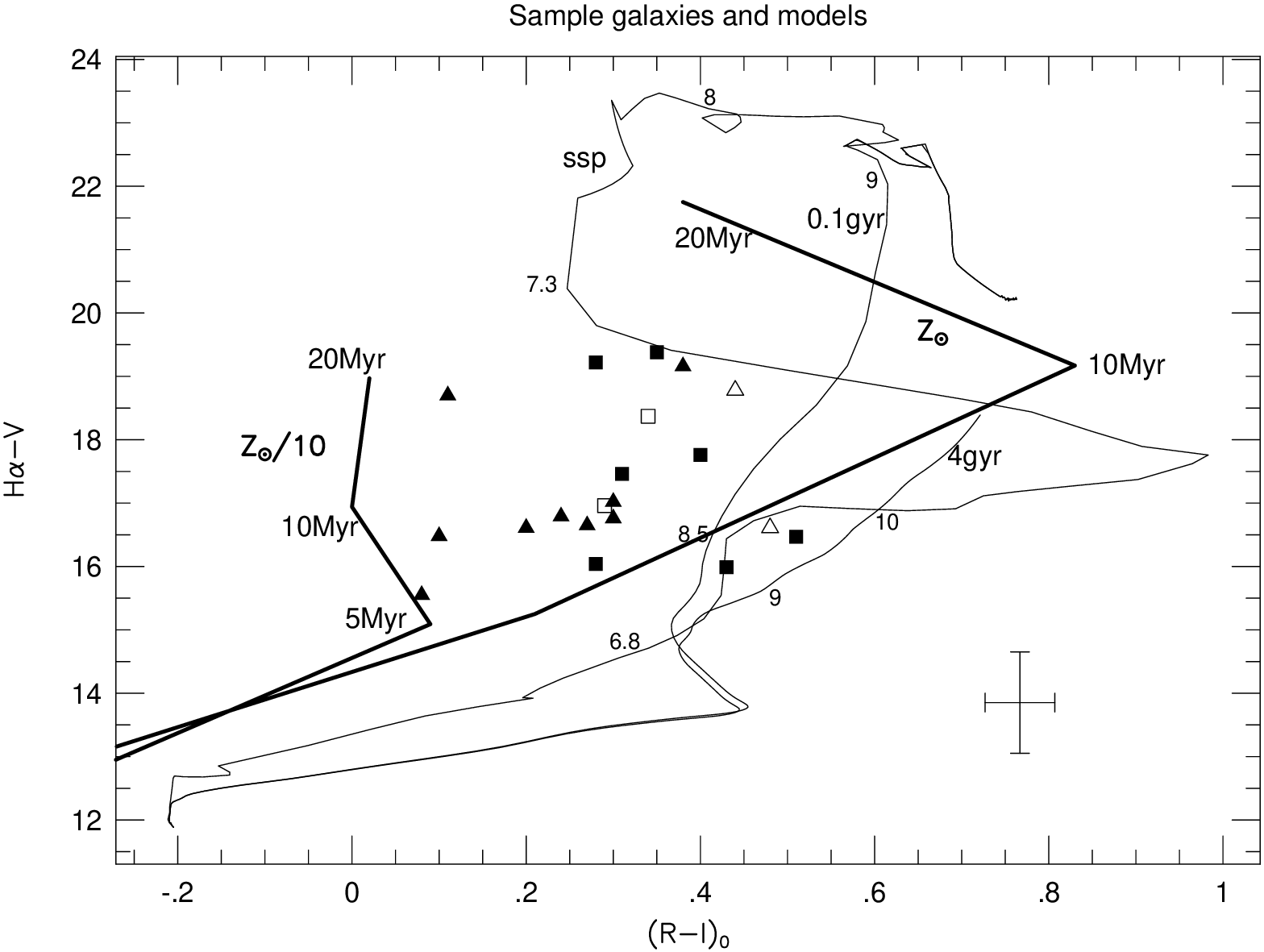}
\caption{\protect \footnotesize{H$\alpha$ --V vs. R--I, with notations the
same as before.}}
\label{fig_HBH3}
\end{figure}

As the nebulosity contribution rapidly decays after $10^7$ years, the
BC93 models describe well the galactic properties after the SFR 
has dropped significantly. 
We should consider, therefore, a combination of the two models, since in most
sample galaxies we have
star-forming regions residing within an older stellar population, and the
models of MHK account for only the first 20 Myr after
the burst. 
Figs.~\ref{fig_HBH1},~\ref{fig_HBH2} and~\ref{fig_HBH3} are the same as
Figs.~\ref{fig_HBV1},~\ref{fig_HBV2} and~\ref{fig_HBV3}, with fewer BC93
models and two models from MHK. These models are for
solar metallicity Z$_\odot$ and Z$_\odot$/10, with IMF slope $\alpha = 2$
 and upper
mass cutoff of 120 M$_\odot$. Other models, with Z$_\odot$, are not 
significantly different from those displayed.
 
While the solar metallicity models are not dramatically different from the 
{\bf ssp} 
BC93 model, the Z$_\odot$/10 model is. The low metallicity model
can explain our results, as it has bluer colors for galaxies with the same
H$\alpha$ line strength. The low metallicity model, together with the 
BC93 models, bracket the observational points. 
We use a simplified two population scenario, considering the two populations
on the diagrams in Figs.~\ref{fig_HBH1},~\ref{fig_HBH2} 
and~\ref{fig_HBH3}, and identifying where their combination falls on the diagrams,
with several relative weightings. Naturally, the result of the combination 
lies on
a monotonous curve joining the two points, each representing one population, 
but the logarithmic scale may be misleading.

\begin{figure}[htbp]
\vspace{14.8cm}
 \includegraphics{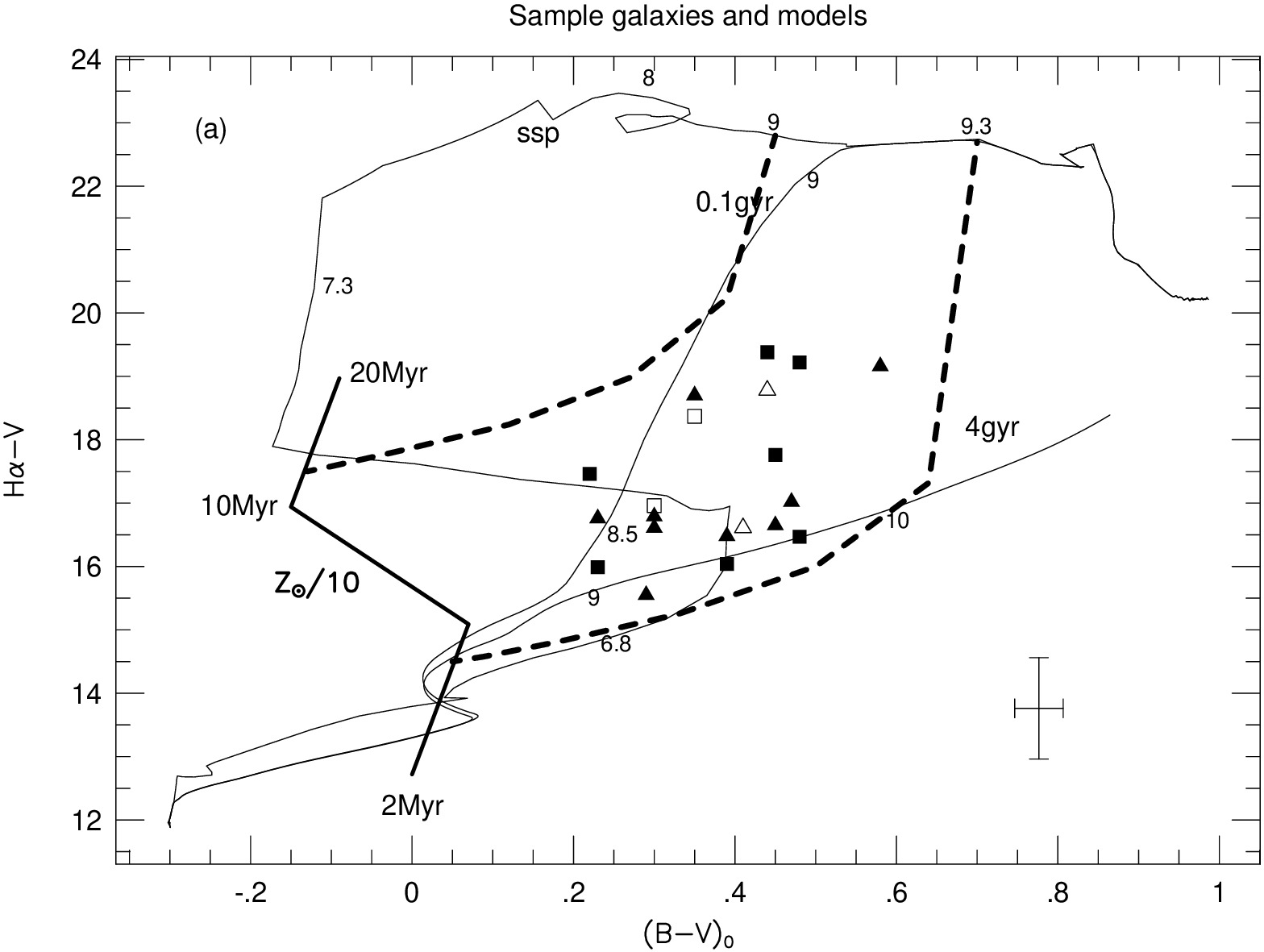}
\includegraphics{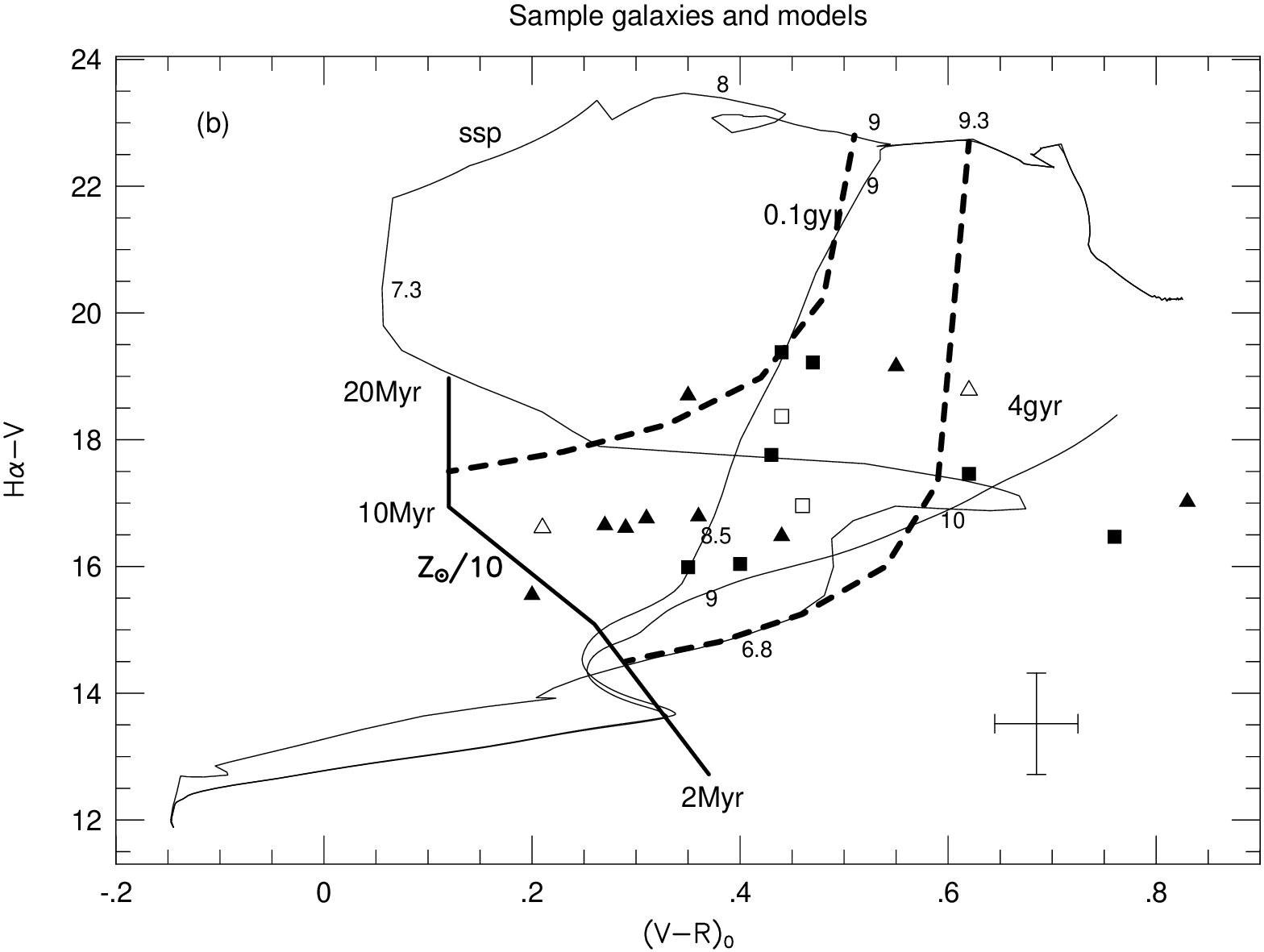}
\caption{\protect \footnotesize{Same as previous figures, with two population 
curves (dashed lines) joining the
low metallicity MHK model with the BC93 ssp model. These curves represent a
mixture with varying ratio of two stellar populations, each located at one
 of the ends of the curves.}}
\label{fig_HX2}
\end{figure}

We calculated combinations of MHK and BC93 models for different V-band
weightings of
1:12.5, 1:3, 1:1, 3:1 and 12.5:1 between the two populations in the
color-color diagrams of Figs.~\ref{fig_HBH1} and~\ref{fig_HBH2}. Joining
the results of these calculations yields a rough curve along which the color
indices of two populations fall in the color-color diagram.
In figures~\ref{fig_HX2}a and~\ref{fig_HX2}b
 we added such `two population curves', connecting
the low metallicity MHK model with ages $\sim$3 and $\sim$13 Myr to the BC93 
{\bf ssp}
model with ages 1 and 2 Gyr. These curves are represented by the thick, dashed 
lines in the figures. 
These lines 
bracket almost all the galaxies in
our sample. The results indicate that the Virgo BCD galaxies may be described as a
combination of a young, low metallicity stellar population of age 3--13 Myr
with an older 1--2 Gyr population, that also originated in a burst of star
formation
(the metallicity factor is less significant for the older stellar population,
therefore it is justified to combine the low metallicity young population
with the old population which has solar metallicity). 

In our derivation,
the populations were weighed by their relative V-band luminosities. 
It is more 
adequate, however, to interpret the relative weight in terms of mass. 
However, due to the rapid evolution of the massive stars, which provide
the bulk of the V-band luminosity at the early stages of a burst, we can
only say that for a galaxy located mid-way between the two populations
only 0.25--5\% of the mass would be in the young burst.
The derived mass ratio of the two populations depends also on
the models used, a fact which, again, demonstrates the inaccuracy of such
a derivation. Thus, the two population curves in Figs.~\ref{fig_HX2} should
be taken as indicative, and it is clear that for most galaxies most
of the mass is in the older stellar population.

It should be noted, regarding this picture, that
 in practice, the older population was not created by a zero duration
burst, but rather by a finite length burst. We may use 
our data to constrain the length of the burst that produced the older stars. 
From Figs.~\ref{fig_HX2}a and~\ref{fig_HX2}b it appears that a slowly
decaying SFR population like the 4Gyr one cannot account for the sample
galaxies. A two-population curve drawn from it to the low metallicity 
model, or to any other model, will not bracket all the galaxies in both 
 diagrams. We may, thus, conclude that the older 
stars in the
sample galaxies are likely to have originated in an exponentially decaying
star formation process of at most $\sim$0.2 Gyr decay time. Similar
behaviour is seen in
the (H$\alpha$--V) vs. (V--I) diagrams. 

One can also note that the locus of the data points seems to be different
regarding the weighting of the young and old populations from one color
to the other. In (B--V) they are more condensed in the center, while in (V--R)
and (R--I) they are more dispersed and seem to be closer to the young,
low metallicity models. We estimate that this might be due to uncertainty
and incompleteness of the models and/or some biases of our broad-band results.
As discussed in paper I and above, we believe the robustness of our
results, and in any case a bias of $\sim 0.1\;mag$ will not alter our 
conclusions concerning the two population scenario.

This scenario, although simple, indicates general star formation
characteristics of the sample galaxies. Obviously, the galaxies are not identical,
neither in their metallicity, nor in their SFR and their dust extinction.
The latter increases the scatter of the observed data points. However,
the conclusion is that some of the galaxies, particularly the bluer ones,
 must have metallicities considerably lower than the solar value.
This finding is not new and is in agreement with data for
other BCD galaxies (e.g., Kunth \& Sargent 1986).

The star formation
process in these galaxies, with no connection to whether they are BCDs or
earlier types, is likely to occur in short duration episodes and the previous 
episodes probably appeared in
the last 1--2 Gyr. Actually, it is more probable that a few episodes, or bursts have occurred 
during the lifetime of the galaxies, rather than just a single one.
This is consistent with other results for blue galaxies, e.g., Huchra {\it
et al.} (1983), Fanelli {\it et al.} (1988) and Deharveng {\it et al.} (1994), which
trace a burst population with an underlying older population.
We can, with a reasonable degree of confidence,
rule out the possibilities that
{\it slowly} decaying past star formation activity is present in 
these dwarf galaxies.

It is worth mentioning that the two-population approach with 
BC93 models alone cannot explain the data points in all
diagrams simultaneously. It is possible, in principle, to explain the 
points in a 
certain color-color diagram by a combination of two populations, but 
combinations of {\it different} populations are needed in other diagrams.
This indicates that no solar metallicity
model can explain the results of our sample as a whole. 

The low metallicity picture can have an alternative explanation, however,
if the LyC photons in the HII regions are heavily extinguished by dust,
or just escape the regions.
This is in contrast to the general picture accepted today,
but, as pointed out by MHK, we do not know exactly by
how much is LyC extinguished. If the original ionizing photons
 are extinguished by an order of magnitude, this will bring the 
observed data
points in accordance with the (two population combination of) solar 
metallicity models. 

This explanation can be tested using the available IRAS
data for some of the sample galaxies. The five H$\alpha$-brightest galaxies
of our sample have IRAS fluxes in both the 60$\; \mu$m and 100$\; \mu$m bands.
We can estimate the SFR in these galaxies from the FIR data, using the
method of Thronson \& Telesco (1986). This method is based on the assumption 
that essentially all of the luminosity of the massive OB stars is absorbed by 
the dust and is reemitted in the infrared. This yields the following relation:
\begin{equation} \label{e_Thro}
SFR = 6.5\times 10^{-10} L_{IR} \ ,
\end{equation}
where the FIR luminosity, $L_{IR}$, which is given in solar units,
can be approximated by:
\begin{equation} \label{e_LIR}
L_{IR} \approx 6\times 10^5 D^2 (2.58 f_{60} + f_{100}) 
\end{equation}
where $D$ is the distance to the object in Mpc, and $f_{60}$ and $f_{100}$
are the two IRAS flux densities, in Jy. 
The resulting SFR is
lower by a factor of two than that derived from H$\alpha$.
This discrepancy can be caused by three
possible reasons:
\begin{enumerate}
\item Our adopted value of 
a factor of two, for the H$\alpha$ flux extinction may be an overestimate.
\item There are discrepancies in the models, such as IMF slope,
high mass cutoff, and estimates of parameters such as the lifetimes of massive
stars and their LyC and total luminosities.
\item Not all the luminosity of the massive stars is absorbed by the dust,
 in particular the LyC radiation is not absorbed by it, as the ambient gas
is a more efficient absorber in these wavelengths, or, the dust does not
reside inside the line-emitting region at all.
\end{enumerate}
It is evident that the first reason alone cannot account for this mismatch.
This is because only the extreme case of no dust at all would bring the
H$\alpha$-estimated SFR in agreement with the FIR-estimated value.
No dust means practically no FIR radiation, thus there should be no
 FIR radiation from
the objects. Adopting a more moderate approach of 
low optical depth of the dust,
will bring the results closer to each other, but never bring them to a complete
agreement. It is, thus, possible that we have overestimated the dust 
extinction of the H$\alpha$ line radiation, though not by a very large extent.
We prefer to retain our adopted 
value of dust extinction for further discussion. 

Therefore the other two reasons
mentioned above must play {\it some} role in changing the estimated SFR value.
Although it is not possible to disentangle the influence of these two
effects, they are both inconsistent with the LyC radiation being heavily
absorbed by the dust. If LyC photons are extinguished by
dust, this would result in a significantly higher FIR radiation
while the H$\alpha$ flux would be further reduced, opposite
to the situation observed here. 
This finding strengthens the scenario that the LyC radiation is {\it not} 
extinguished significantly in the cores of HII regions.

The comparison between the FIR and H$\alpha$ radiation sheds light
upon the optical depth of the hydrogen clouds and their covering factor. 
If the HII regions are
not ionization bound, this will cause a reduction of the H$\alpha$ flux
and may enhance the FIR emission from the galaxies,
because the LyC photons that escape the HII regions are likely to be absorbed 
by more distant diffuse dust.
This dust will re-radiate
its energy in the FIR. Again, the effect is opposite to the current situation,
a fact which strengthens our assumption that nearly all the LyC flux is
absorbed by the hydrogen.

It is clear that different IMFs influence
 the observational parameters less than the metallicity and age of a
stellar population, but with the dispersion of our data points, and with many
additional parameters, it is practically impossible to determine the IMF of the
galaxies. Other factors, like gas recycling
in the model galaxy, are even less significant and cannot be tested here.

\begin{figure}[htbp]
\vspace{7.5cm}
\includegraphics{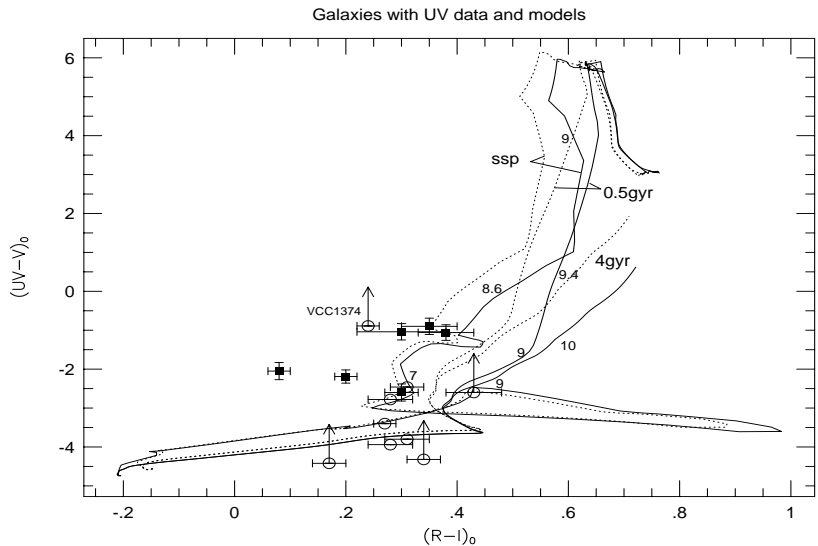}
\caption{\protect \footnotesize{UV--V vs. R--I for the sample galaxies,
 together
with models from BC93. Galaxies with significant UV results are 
marked by filled squares, and those with UV upper limits as open circles.
 For clarity, not all upper limit arrows are drawn.
For the theoretical UV we use the data of HST-UV14 
filter given by BC93.}}
\label{fig_UVI}
\end{figure}

In addition to the H$\alpha$ data, it is important to consider the 
UV data available for some of our galaxies. The UV radiation (at
$\lambda \geq$ 1300\AA, in our case) is also emitted mostly
 by the young massive stars and is more sensitive to the star formation 
properties of
the galaxies than the optical colors. We calculated, for the galaxies 
which have UV data, the
\mbox{(UV--V)} color (UV is the monochromatic magnitude
at $\sim$1650\AA, as discussed in paper I). We corrected these values
for dust extinction, according to the value 
 in section~\ref{sec_OBS}, and compare them to 
the "14--V" color, given by BC93 for the HST-UV14 filter. 
In Fig.~\ref{fig_UVI} the galaxies are plotted in the (UV--V) vs. (R--I)
plane, together with BC93 models. The galaxy VCC1374, which has a faint UV
upper limit, is bright in H$\alpha$ and in the optical. 

The UV results are consistent with
our previous interpretations for the star formation properties of the galaxies.
The same trend is observed as in the 
H$\alpha$ data for the three colors B--V, V--R and R--I, and only the latter
is displayed here with the UV data. Unfortunately, only few  
galaxies have significant UV fluxes from IUE or from FAUST observations.

Instead of a color-color diagram, it may be useful to consider the entire SEDs
of the galaxies, as derived in paper I. Since our derived
SEDs are very sparse because of the broad-band colors,
 they cannot be tested against detailed spectra, but maybe
compared, in general, with typical SEDs of different stellar populations.
Using the set of spectra published by BC93, it can be noted that the galaxies
VCC144, VCC1725 and VCC1791, which are the UV and H$\alpha$-brightest in our 
sample, may correspond to a young stellar population $\sim$0.1--0.3 Gyr old,
regardless of the details of the star formation history. The situation of
VCC10, VCC22 and VCC24 is less clear, however, since their SEDs are relatively 
flat in the optical, and may correspond to an age of 4--13 Gyr.

%
The question that arises at this stage concerns the difference among the 
various sample galaxies in terms of their star formation.
In light of the 
interpretation of the results discussed above, of a very young, low
metallicity, stellar population residing within (probably a larger 
region which contains)
an older stellar population, we conclude that the difference lies in the
relative weight of the two populations (in terms of their V magnitude,
in our method of derivation). In the bluer and H$\alpha$-brighter 
galaxies, the young stars currently forming dominate the entire
spectrum of the galaxy (although not necessarily dominating the mass), while in 
the red, H$\alpha$-faint objects, the star
formation is relatively weak and the galactic colors are determined more by
the older stellar population. As discussed above, the older stellar
population is believed to have originated in one or more $\sim$instantaneous
 bursts of star formation. 

The arguments presented above lead to the conclusion that all
the sample galaxies, classified as BCDs or Im III--IV, or even
earlier types, are basically
of the same type, in which the star formation appears in bursts, but which are
observed at different epochs of their lifetimes - some during the burst
and others at a relatively quiescent epoch. In this context, galaxies like
VCC1725, VCC1791 and VCC144 are examples of strong star-forming galaxies, in
which the underlying older stellar population does not contribute much to
the SED. In VCC144 there is hardly a sign of the older stellar population,
and we conclude that this galaxy may be experiencing its first star formation 
burst (Brosch {\it et al.} 1997).
 On the other hand, VCC10,
VCC22 and VCC24 are examples of galaxies in which the older stellar population
dominates the SED.

Although this conclusion is not new, it is nicely supported by our results. If
we could identify all the galaxies of this same type, maybe all the 
low-metallicity dwarfs, we would be able to conclude how often a burst takes
place in these galaxies. To be more precise, we would be able to determine the 
duration
of the burst period relative to the quiescent period from the 
fraction of high SFR galaxies observed in the entire sample. 

\subsection{Life expectancy of the galaxies}
 \label{sub_lif}

A starburst galaxy is one in which the current SFR
is very high, relative to its
average past SFR. Another indicator of a starburst galaxy
is its gas consumption time. This is the time by which all the hydrogen 
available for star formation will be exhausted, assuming the current SFR
is maintained. If this consumption time is considerably shorter than one
Hubble time, the galaxy is probably experiencing a starburst.
 The current high SFR must then be a
transient situation, after which the galaxy is expected to return to a
`normal life', as far as the SFR is concerned.

The gas consumption time, or "Roberts time" (Sandage 1986),
is given by: \linebreak \mbox{$\tau_R = \frac{M_{gas}}{SFR}$,}
 where $M_{gas}$ is the total mass of interstellar gas in the
galaxy. For the derivation of Roberts times of the sample galaxies, we used
 their HI mass from {\mbox paper I.}
 The total hydrogen
mass may be at most $\sim$20\% higher than the HI mass, as
in late-type galaxies the H$_2$ is only a small part of the total 
hydrogen content (Young \& Knezek 1989).

%
\begin{table}[htbp]
\vspace{9.3cm}
\includegraphics{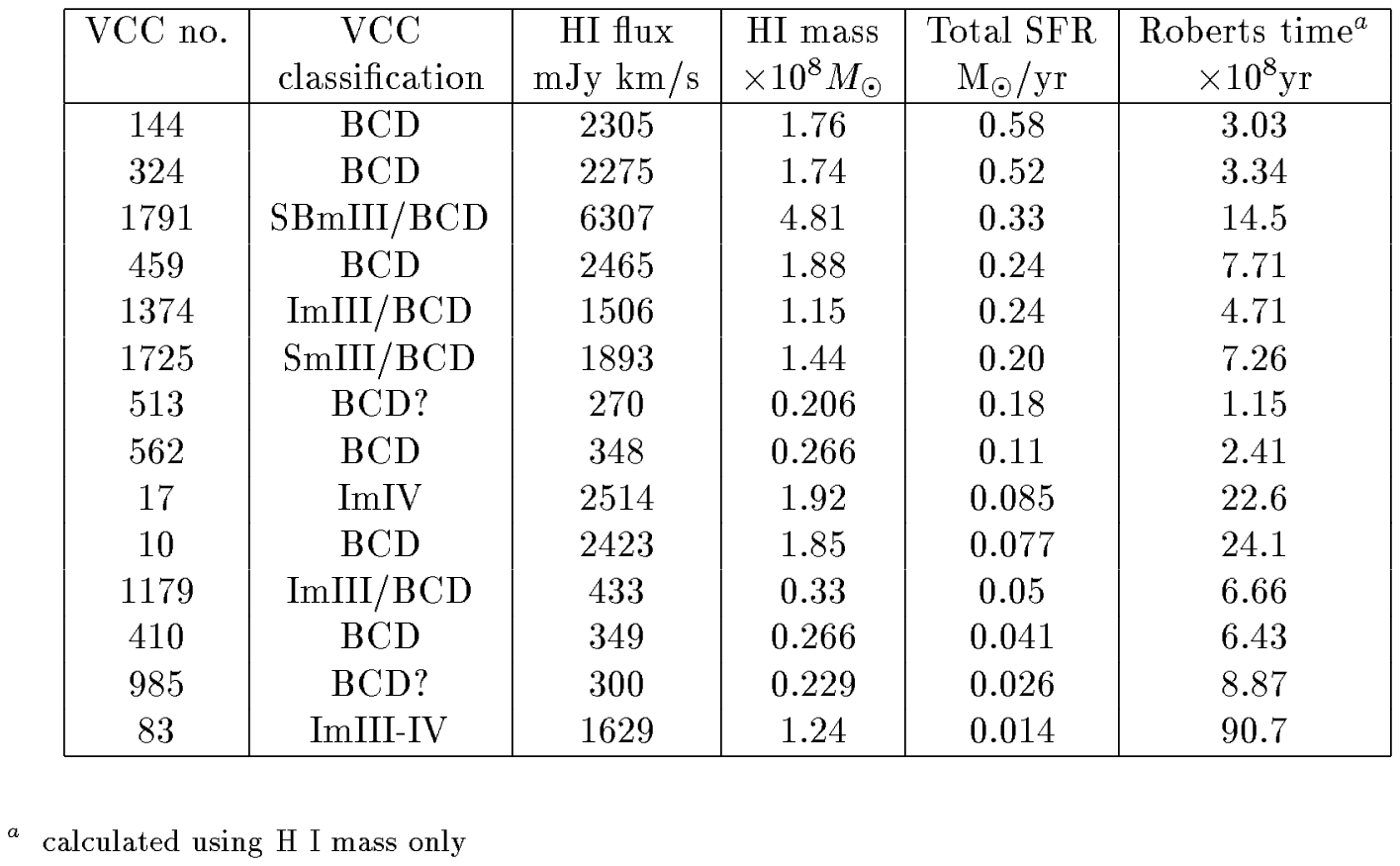}
\caption{\protect \footnotesize{HI properties of VCC galaxies.}} 
\label{tab_HTr}

\end{table}

\begin{figure}[htbp]
\vspace{7.5cm}
\includegraphics{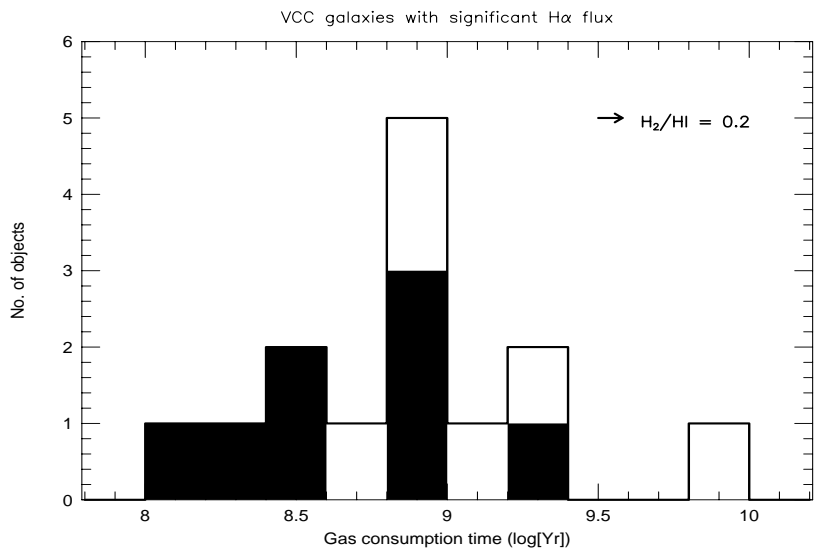}
\caption{\protect \footnotesize{Histogram of the gas consumption times for
 VCC galaxies. The entire
sample is shown, and BCDs are
in black. The extension of the gas consumption time, due to the contribution of
H$_2$ at 20\% of the HI, is indicated by an arrow. This represents an upper
limit to the H$_2$ contribution (see text).}}
\label{fig_Rti}
\end{figure}

In Table~\ref{tab_HTr} we tabulate the HI properties of the sample galaxies.
 The distribution of Roberts times is presented in 
Fig.~\ref{fig_Rti}, where the galaxies classified as "BCD" or "BCD?" in  
VCC are marked in black. Clearly, many of the galaxies 
are starbursts, 10 out of 14 having a
gas consumption time smaller than 1 Gyr. Even when including the maximally
possible H$_2$ mass
contribution, the times are rather short.

The Roberts' times are not the real life expectancies
of the galaxies, even with the present SFR, because gas 
recycling 
may extend considerably the star formation. The factor by which the
recycling prolongs the life expectancy of the galaxies depends basically
on three parameters:
\begin{enumerate}
\item The fraction of mass returned to the interstellar medium by a single 
age stellar population,
\item The efficiency of converting the available gas to stars, and the 
dependence of the SFR on the gas density,
\item The amount of hydrogen available originally.
\end{enumerate}

The gas return fraction strongly depends on details of the IMF,
metallicity etc. Thus, it is difficult to assess
the exact effect of recycling on the star formation of these galaxies. 
For disk galaxies, Kennicutt {\it et al.} (1994) obtained a range of 
\mbox{1.5--4}
for this prolonging factor.
If we adopt these values here we obtain
a typical expected
lifetime shorter than 2--4 Gyr, which is relatively short.
In irregular galaxies, in contrast to large disk galaxies, a
large fraction of the HI extends beyound the optical size, where the
star formation activity is taking place. This will reduce the future lifetime of
the galaxies even further, provided that no star formation will start outside
the visible optical radius. Thus, the uncertainty of the expected future
lifetimes of the galaxies in our sample is rather large, but still there is a strong
 indication of a starburst ongoing in these galaxies.

This result supports our picture of objects in which the star formation 
appears in bursts. The real future lifetime of the galaxies depends on the
duration of the starbursts and the time interval between bursts. This
 question is still open, although
our interpretation indicates that previous bursts probably happened
in most of the galaxies during the last Gyr.

\subsection{Star formation mechanisms}
 \label{sub_sfme}

The question that is raised by the above discussions is how do the different 
star formation inducing mechanisms govern the
star formation in the late-type dwarfs of our sample. Any answer to
this question can only be found by a large survey, which
encompasses {\it all} galactic types.
 However, as derived here,
once the star formation activity is triggered, its
intensity, as manifested by the SFR/area, appears similar in small
and large galaxies. 
This indicates that
small-scale mechanisms, believed to dominate in dwarf
galaxies, are as efficient as the
large-scale mechanisms acting in large galaxies.

 A note should be made, however, concerning the comparison of the
SFR/area in our galaxies with that in large disk galaxies, where the intensive
star formation process is confined to a few hundred pc. thick
disk. This is because dwarf galaxies are generally amorphous and
diskless, thus their observed SFR/area is actually the
`column density' of the SFR. It would be more appropriate to characterize
the intensity of the star formation process in these galaxies by
spatial density rather than by surface density, which, of course, is not 
possible here. One should note, however, that the actual thickness of a dwarf galaxy
may not be too different from that of a disk in a large galaxy, thus
the comparison may not be totally out of order.

Another phenomenon, believed to trigger star formation in
clusters of galaxies, is the gravitational tide from a neighboring 
galaxy, which is
strongest near the cluster core where galactic encounters are more frequent. 
Tidal forces can trigger
star formation also in field galaxies with close companions - many
field starburst galaxies are found to be interacting with companions.
It is, thus, natural to expect a relation between the star formation
properties of the sample galaxies and their distance from the center
of the Virgo cluster. The star formation is expected to be more
pronounced in galaxies that are close to the center. 
However, no significant correlation was found between the
 Virgocentric distance and any of the star formation
parameters.  If anything, the SFR seems to increase with increasing
Virgocentric distance, while the SFR/area does not show any
correlation. The SFR/area also does not show any correlation with the recession
velocity of the galaxies.
This implies that tidal forces near the core of the Virgo cluster
are not significant relative to other
mechanisms triggering star formation in dwarf galaxies. 

An additional cluster effect, which may influence the star formation
properties,
 is the stripping of interstellar matter from galaxies that pass near the
 cluster core (e.g., Haynes {\it et al.} 1990).  
Galaxies near a
cluster core are found to have less HI than galaxies further away from
the core. The high HI galaxies of our
sample, which are more than 6$^\circ$ away from the center, have higher
HI flux, on average, than those closer than 6$^\circ$. This can be seen also
in Fig. 2 of paper 1, where the low HI galaxies are 
distributed
more or less evenly, while the high HI galaxies appear only in the outer
parts of the cluster. This effect is also seen using a larger sample of Virgo
dwarfs from Hoffman {\it et al.} (1987, 1989).

The gas-stripping
phenomenon may affect the star formation in the galaxies studied here
 in an opposite
direction to that of tidal forces discussed above, the
amount of HI available to form stars being less near the core than away from
it. We believe
it would be an overinterpretation of the data, however, to say that the
apparent increase in SFR with Virgocentric distance is due to this effect.

\section{Conclusion} \label{sec_conc}

The main goal of this study is to investigate mechanisms that govern 
star formation processes in galaxies and their dependence on various
galactic parameters.
The intention in focusing on our sample of late-type dwarfs in the Virgo 
cluster was to exclude some of these mechanisms
thought to be responsible for 
star formation in large galaxies. 
In addition, we concentrate on
cluster members to test for effects of the cluster 
environment on the star formation properties of the galaxies.

The observational data used to evaluate the star formation
parameters, such as the SFR, IMF, and star formation history, are affected
primarily by the internal dust extinction in the galaxies. This
depends on the amount and distribution of dust in the galaxy.
We adopted general correction parameters for the entire sample
 to account for the effects of dust. 

Using a data base consisting of a number of broad band colors and
H$\alpha$ observations it is possible to track
the ongoing star formation process, as well as the star formation history of
the sample. This is done for both low HI and high HI subsamples, which 
enables one to
check the dependence of these parameters on the neutral hydrogen content of
the galaxies.
In all cases, 
no significant dependence of the star 
formation
properties on the HI content was found. This may be explained by the following 
argument: the differences among the 
various galaxies in our sample lie in the relative weight of the flux 
originating from their `current' and `previous' star
formation episodes. Since the hydrogen is depleted 
during each such burst of star formation, its current amount depends on the 
number of bursts that
occurred in the past, as well as on the original amount present. 
 The strength of the current starburst apparently does 
not depend on this star formation history and, thus, cannot be correlated with 
the neutral hydrogen content.

Considering the entire set of observations, together with various 
population synthesis models, a star formation scenario can be sketched
 for the sample
galaxies. The process of star formation in late-type
dwarf galaxies takes place in short bursts, presumably much shorter than 1Gyr
(10--100 Myr). The past burst(s) probably occurred within the
last $\sim$Gyr. Redder galaxies can be fitted with a single,
longer burst with this decay time (1 Gyr), but the observational data of
the entire sample can only
be explained in terms of a series of short bursts. In this picture,
 the difference among  
galaxies lies in the relative weight of
the starburst and the older population. This depends on the
age and size of the current burst and of former bursts, therefore
 the dwarf galaxies
may be seen as a single type, which are observed
at different epochs of their evolution.

In addition, the galaxies appear to have low metal abundance, mainly
from their blue R--I color. This is in
agreement with previous data for irregular dwarf
galaxies, known to have typically low metallicities 
(e.g., Kunth \& Sargent 1986).
A low metallicity of a galaxy indicates its young age, since the amount of
heavy elements produced by massive stars increases during the
lifetime of the galaxy, which
implies that some of our sample galaxies are genuinely young. 
This also is not a
new finding (see Gondhalekar {\it et al.} 1984).
A development trend 
 can be sketched, in which the galaxies undergo a series of
starbursts, and their metallicity increases from one burst to the
other. Since the HII regions in many of the galaxies occupy a large
fraction of the galactic volume, it is possible that each burst
 changes significantly the total galactic metal abundance. The relation
between the fraction
of the galactic surface covered by H II regions and star formation properties
will be discussed in a subsequent paper.

We have tested the cluster influence on the sample galaxies' star formation properties.
No correlation was found between the Virgocentric distande of the galaxies
and any of their star formation properties. This may manifest the low significance
of tidal forces when acting on dwarf galaxies. It can be understood as
these galaxies are small in size and, thus, a gradient in an external gravitational
field induced by another galaxy may not cause a great difference from side to side
relative to the galaxy's own gravitational well. This finding is, therefore, not
surprising.

A remarkable galaxy in our sample is VCC144. It has strong SFR
and an exceptional SFR/area, compared with the other galaxies. It is also the
only object in which the burst population is believed to dominate in 
luminosity and in mass over older populations, if any.
However,
it does not show any special behavior in other parameters such as the
HI flux, velocity dispersion, infrared flux, or distance from cluster
center, in which it seems a `normal' object in the sample. It is the
most condensed object of our sample, in terms of star formation, but with no 
other
peculiarities. It is possible that this object is experiencing 
its very first burst of star formation, which
 indicates that galaxies are still being formed these days in the Virgo
cluster. 

To conclude - the star formation in late-type dwarf galaxies
 in the Virgo cluster
occurs probably in bursts. The bursts do not seem to depend on galactic 
history or on cluster environment. The details of the IMF of the
burst population cannot be determined from the data collected in this study,
 but a general scenario of the
galactic evolution is sketched. During their lifetime, the galaxies
evolve from late-type to earlier type, their metallicity increases, and
they become redder objects in the optical. 

\section*{Acknowledgments}

We would like to thank the referee for the constructive remarks.

Multi-spectral observations at the Wise Observatory are partly
 supported by a Center of 
Excellence Grant from the Israel Academy of Sciences.
UV studies at the Wise Observatory are supported by special grants from
the Ministry of Science and Arts, through the Israel Space Agency, to
develop TAUVEX, a UV space imaging experiment, and by the Austrian Friends
of Tel Aviv University. 

EA acknowledges a grant from "The Fund for the
Encouragement of Research" Histadrut- The General federation of Labour in
Israel.
NB acknowledges the hospitality of Prab Gondhalekar
and of the IRAS Postmission Analysis Group at RAL, as well as IRAS Faint
Source catalog searches by Rob Assendorp.

We thank Stuart Bowyer and Tim Sasseen from the Space Sciences Laboratory, 
Berkeley, University of California,
 for kindly providing the FAUST images of the Virgo cluster.

\section*{References}
\begin{description}

\item{Allen, C.W. 1973, {\it Astrophysical Quantities}, The Athlone Press,
University of London.} 

\item{Brosch, N., Almoznino, E. \& Hoffman, G.L. 1997, Astron. Astrophys. {\it in press}.}

\item{Bruzual, G.A., \& Charlot, S. 1993, Astrophys. J. {\bf 405}, 538 (BC93).}
 
\item{Buat, V., Deharveng, J.M. \& Donas, J. 1989, Astron. Astrophys. 
{\bf 223}, 42.}
 
\item{Calzetti, D., Kinney, A.L. \& Storchi-Bergmann, T. 1994, Astrophys. J. 
{\bf 429}, 582.}
 
\item{Deharveng, J. M., Albrecht, R., Barbieri, C.,
 Blades, J. C., Boksenberg, A., Crane, P.,
 Disney, M. J., Jakobsen, P., Kamperman, T. M.,
 King, I. R., Macchetto, F., Mackay, C. D.,
 Paresce, F., Weigelt, G., Baxter, D., Greenfield, P., Jedrzejewski, R., 
Nota, A., Sparks, W. B. 1994, Astron. Astrophys. {\bf 288}, 413.}

\item{Fanelli, M. N., O'Connell, R. W., Thuan, T. X. 1988, Astrophys. J. {\bf 334}, 665.}
 
\item{Fernie, J.D. 1983, PASP {\bf 95}, 782.} 

\item{Gallagher, J.S., Hunter, D.A. \& Tutukov, A.V. 1984, Astrophys. J. {\bf 284}, 544 (GHT).}

\item{Gondhalekar, P.M., Morgan, D.H., Dopita, M. \& Phillips, A.P. 1984, Mon. Not. R. astr. Soc. 
{\bf 209}, 59.}

\item{Haynes, M.P., Herter, T., Barton, A.S. \& Benensohn, J.S. 1990, 
Astron.J. \mbox{{\bf 99}, 1740.}} 

\item{Hoffman, G.L., Helou, G., Salpeter, E.E., Glosson, J. \& Sandage, A.
 1987,  Astrophys. J. Suppl. {\bf 63}, 247.}

\item {Hoffman, G.L., Williams, H.L., Salpeter, E.E., Sandage, A. \&
Binggeli, B. 1989, Astrophys. J. Suppl. {\bf 71}, 701.}

\item{Huchra, J. P., Geller, M. J., Gallagher, J., Hunter, D., Hartmann, L., Fabbiano, G., Aaronson, M. 1983, Astrophys. J. {\bf 274}, 125.}

\item{Kennicutt, R.C. 1983, Astrophys. J. {\bf 272}, 54 (K83).}
 
\item{Kennicutt, R.C. \& Kent, S.M. 1983, Astrophys. J. {\bf 88}, 1094.}
 
\item{Kennicutt, R.C. 1989, {\it Large Scale Star Formation \& the Interstellar
 Medium}, in "The Interstellar Medium in External Galaxies". ed.
 H.A.Thronson \& J.M.Shull.}
  
\item{Kennicutt, R.C., Tamblyn, P. \& Congdon, C.W. 1994,  Astrophys. J. 
{\bf 435}, 22.}
 
\item{Kunth, D. \& Sargent, W.L.W. 1986, Astrophys. J. {\bf 300}, 496.}

\item{Larson, R.B. 1987, {\it Star Formation Rates \& Starbursts}, in "Starbursts \&
 Galaxy Evolution". ed. T.X.Thuan, T.Montmerle \& J.T.T.Van
 (Edition Frontieres, Gif sur Yvette - FRANCE).}

\item{Mas-Hesse, J.M. \& Kunth, D. 1991, Astron. Astrophys. Suppl. {\bf 88}, 399.}

\item{Miller, G.E. \& Scalo, J.M. 1979, Astrophys. J. Suppl. {\bf 41}, 513 
(MS79).}
 
\item{Osterbrock, D.E. 1989, {\it Astrophysics of Gaseous Nebulae \& Active Galactic
 Nuclei}, (Mill Valley, CA: University Science Books).}
 
\item{Pogge, R.W. \& Eskridge, P.B. 1987, Astrophys. J. {\bf 93}, 291 (PE87).}

\item{Salpeter, E.E. 1955, Astrophys. J. {\bf 121}, 161 (S55).}

\item{Sandage, A. 1986, Astron. Astrophys. {\bf 161}, 89.}

\item{Savage, B.D. \& Mathis, J.S. 1979, Ann. Rev. Astron. Astrophys. (SM).}
 
\item{Scalo, J.M. 1986, Fundamentals of Cosmic Physics {\bf 11}, 1 (S86).}

\item{Schmidt, M. 1959, Astrophys. J. {\bf 129}, 243.}

\item{Shields, G.A. 1990, Ann. Rev. Astron. Astrophys. {\bf 28}, 525.}

\item{Terlevich, R. \& Melnick, J. 1983, ESO preprint no. 264.}
 
\item{Thronson, H.A. \& Telesco, C.M. 1986, Astrophys. J. {\bf 311}, 98.}

\item{Young, J.S. \& Knezek, P.M. 1989, Astrophys. J. Lett. {\bf 347}, L55.}

\end{description}


\end{document}